\documentclass[%
    twocolumn,
    superscriptaddress,
    bitnotes,
    amsmath, amssymb, 
    aps, 
    prx,
    floatfix
]{revtex4-2}

\usepackage{changes}
\usepackage{graphicx}       
\usepackage{dcolumn}        
\usepackage{bm}             
\usepackage{color}
\usepackage{subfigure}
\usepackage{appendix}
\usepackage[utf8]{inputenc}
\usepackage{hyperref}
\hypersetup{
    colorlinks=true,
    linkcolor=blue,
    filecolor=magenta,      
    urlcolor=cyan,
    pdfpagemode=FullScreen,
}

\definecolor{royalblue}{rgb}{0.25, 0.41, 0.88}
\definecolor{royal_purple}{RGB}{120, 48, 212}
\definecolor{darkgreen}{RGB}{1,113,0}

\begin{document}
\title{Investigating a Quantum-Inspired Method for Quantum Dynamics}

\author{Bo Xiao}
\email{xiaob@ornl.gov}
\affiliation{Center for Computational Quantum Physics, Flatiron Institute, 162 Fifth Avenue, New York, New York, 10010, USA}
\affiliation{Materials Science and Technology Division, Oak Ridge National Laboratory, Oak Ridge, Tennessee 37831, USA}
\affiliation{Quantum Science Center, Oak Ridge, Tennessee 37831, USA}

\author{Benedikt Kloss}
\affiliation{Center for Computational Quantum Physics, Flatiron Institute, 162 Fifth Avenue, New York, New York, 10010, USA}
\affiliation{NVIDIA Corp, Santa Clara, CA, USA}

\author{E. Miles Stoudenmire}
\email{mstoudenmire@flatironinstitute.org}
\affiliation{Center for Computational Quantum Physics, Flatiron Institute, 162 Fifth Avenue, New York, New York, 10010, USA}

\date{\today}
\begin{abstract}
Building on recent advances in quantum algorithms which measure and reuse qubits and in efficient classical simulation leveraging projective measurements, we extend these frameworks to real-time dynamics of quantum many-body systems undergoing discrete-time and continuous-time Hamiltonian evolution, and find improvements that significantly reduce sampling overhead. The approach exploits causal light-cone structure by interleaving time and space evolution and applying projective measurements as soon as local subsystems reach the target physical time, suppressing entanglement growth. Comparing to time-evolving block decimation, the method reaches longer times per sample for the same resources. We also gain the ability to study dynamics of entanglement that would be occurring on quantum hardware when following similar protocols, such as the holographic quantum dynamics simulation framework.
We show how to efficiently obtain local observables as well as equal-time and time-dependent correlation functions. Our findings show how optimizations for quantum hardware can benefit classical tensor network simulations and how such classical methods can yield insights into the utility of quantum simulations. 
\end{abstract}

\maketitle

\section{Introduction} \label{sec:intro} 

Simulating the real-time dynamics of quantum many-body systems remains a central challenge in physics, with profound implications for quantum computing. Quantum dynamics simulations enable the computation of transport coefficients and excitation spectra, with applications to materials science and energy technologies, and the bridging of theory and experiment. Such simulations also give understanding of fundamental questions in non-equilibrium
physics, such as the nature of exotic quantum states in systems driven out of equilibrium. An exciting example is the observation of transient high-temperature superconductivity signatures after excitation with a strong laser pulse~\cite{Fausti2011, Budden2021, Cavalleri2018}.

There are well-developed classical computational methods for quantum systems in equilibrium, but more efficient methods are needed for quantum dynamics. Classical simulations routinely encounter exponential costs, for example quantum Monte Carlo generically encounters the sign problem and tensor network methods are strongly limited by growth of entanglement in time.
Recent algorithmic advances have significantly improved the reach of tensor network simulations for special settings, such as 
quantum dot or impurity-model systems~\cite{Nunez2022, Shinaoka2023, Erpenbeck2023, Cao2024, Grundner2024}, transport at high or infinite temperature~\cite{Nahum2018, Keyserlingk2018, Parker2019, Rakovszky2022}, or low-order correlation functions~\cite{Cao2024, Grundner2024,Cha2025}. Frameworks such as influence functional or process tensor approximations of environments also show significant promise for tackling broad classes of dynamics~\cite{Keeling2025Process}, including for higher-dimensional systems~\cite{ParkBP}.
But many challenges remain for simulations of quantum dynamics. 

The present limitations of classical methods have been a key motivation for developing programmable quantum simulators and quantum computing platforms, including cold-atom arrays and superconducting qubit systems. These platforms offer two complementary approaches: analog quantum simulation, where the physical dynamics of the device realize the target Hamiltonian, and digital quantum simulation, where the dynamics are encoded into a quantum circuit and implemented through gate-based operations.

Yet the capabilities of tensor network methods and near-term quantum hardware are actually remarkably similar: for example, in both settings one can apply some number of arbitrary gates and perform projective measurements. The primary distinction lies in the nature of their respective limitations. Tensor networks are limited to moderately entangled states, while present-day quantum hardware is limited by gate fidelity and coherence times, which restrict the circuit depth before noise dominates. Otherwise, developments in quantum algorithms can be readily transferred to tensor network algorithms and vice versa. For example, proposals for simulating differential equations on quantum hardware have been successfully ``ported'' to tensor networks, leading to quantum-inspired classical algorithms with improved scaling~\cite{KhoromskijOseledets2010,khoromskij2014tensor,Lubasch2018,GarciaRipoll2021quantuminspired,gourianov2022quantum,Shinaoka2023}.

Motivated by the limited qubit counts in near-term devices, ``holographic'' algorithms for quantum simulation tasks have been proposed which reuse a finite number of qubits, by interspersing qubit measurement and reset, to simulate large or even half-infinite systems on quantum hardware~\cite{Foss-Feig2021,Chertkov2022}. A notable example is the holographic quantum dynamics simulation (holoQUADS)~\cite{Chertkov2022}, which leverages these ideas to emulate the quantum dynamics of a Floquet system. This naturally raises the question of whether such holographic or mid-circuit measurement ideas can also benefit classical tensor network simulations, which are not constrained by Hilbert space size or number of qubits but instead constrained by entanglement. Indeed, a closely related idea, termed space-evolving block decimation (SEBD) was developed that also takes advantage of qubit measurement and reset, but in a classical simulation context~\cite{Napp2022}. Follow up work inspired by SEBD has included using measurements to simulate physical noise and assess the resulting simulation complexity~\cite{Cheng2023,Cichy2025}, and works optimizing measurement trajectories to improve classical simulability of circuits~\cite{Chen_Optimized_2024,Granet2025}.

The original SEBD study focused on the task of sampling the output of shallow quantum circuits~\cite{Napp2022}. In this paper, we explore whether interspersing real-time evolution and projective measurements yields benefits for simulating the Hamiltonian evolution of closed quantum systems. We also investigate whether important observables, such as time-dependent correlation functions, can be estimated efficiently in the SEBD framework.

More specifically, our goal is to quantify how interspersing projective measurements with light-cone-based time evolution impacts the efficiency of classical simulations of real-time dynamics. Projective measurements suppress entanglement growth in the time-evolved state, which should lead to substantially smaller tensor network bond dimensions. This benefit trades off against the overhead of sampling many random measurement outcomes and the resulting statistical uncertainty. To address this issue, we introduce an entangled measurement (EM) estimator that takes advantage of the availability of a many-body wavefunction at each step,  reducing sample complexity for low-order observables.

We also find the following notable results for extending SEBD- and holoQUADS-like ideas to the simulation of Hamiltonian dynamics:
\begin{itemize}
    \item{We demonstrate that the SEBD framework can be systematically extended from discrete-time regime to the continuous-time regime, maintaining efficiency.}
    \item{Local observables and equal-time correlators at arbitrary distances can be measured with high accuracy and sample efficiency in this framework, especially when using the entanglement measurement framework we describe in Section~\ref{sec:correlation_functions}.}
    \item{Unequal-time (time-dependent) correlators can be measured efficiently within this framework, enabling access to dynamical response functions and spectral quantities.}
    \item{Entanglement is substantially reduced by projective measurements, resulting in potential computational ``wall-time'' advantages over traditional matrix product state (MPS)-based time evolution such as time-evolution block decimation (TEBD), with the advantage growing for longer times. The computational advantage likely requires parallel computing resources.}
\end{itemize}
Our framework also provides quantitative insight into the entanglement dynamics that would emerge in holographic quantum simulations on trapped-ion processors, offering a classical benchmark understanding measurement-induced disentanglement on hardware when using holographic protocols.

This paper is organized as follows. Sec.~\ref{sec:models_methods} introduces the algorithms we use, which interleave real-time evolution and projective measurements along light cones, and outlines the entangled measurement approach for enhancing sampling efficiency. Sec.~\ref{sec:entanglement_dynamics} investigates the entanglement dynamics under SEBD, quantifies its computational advantage over TEBD, and demonstrates its effectiveness even in the continuous-time limit. Sec.~\ref{sec:correlation_functions} describes efficient protocols for computing physical observables---including equal-time and unequal-time correlation functions and their interplay with underlying entanglement dynamics. Finally, in Sec.~\ref{sec:summary}, we highlight implications for quantum experiments and situate SEBD within the broader context of recent advances for extending classical simulations of quantum dynamics.

\begin{figure*}[t]
    \centering 
    \includegraphics[width=1\textwidth]{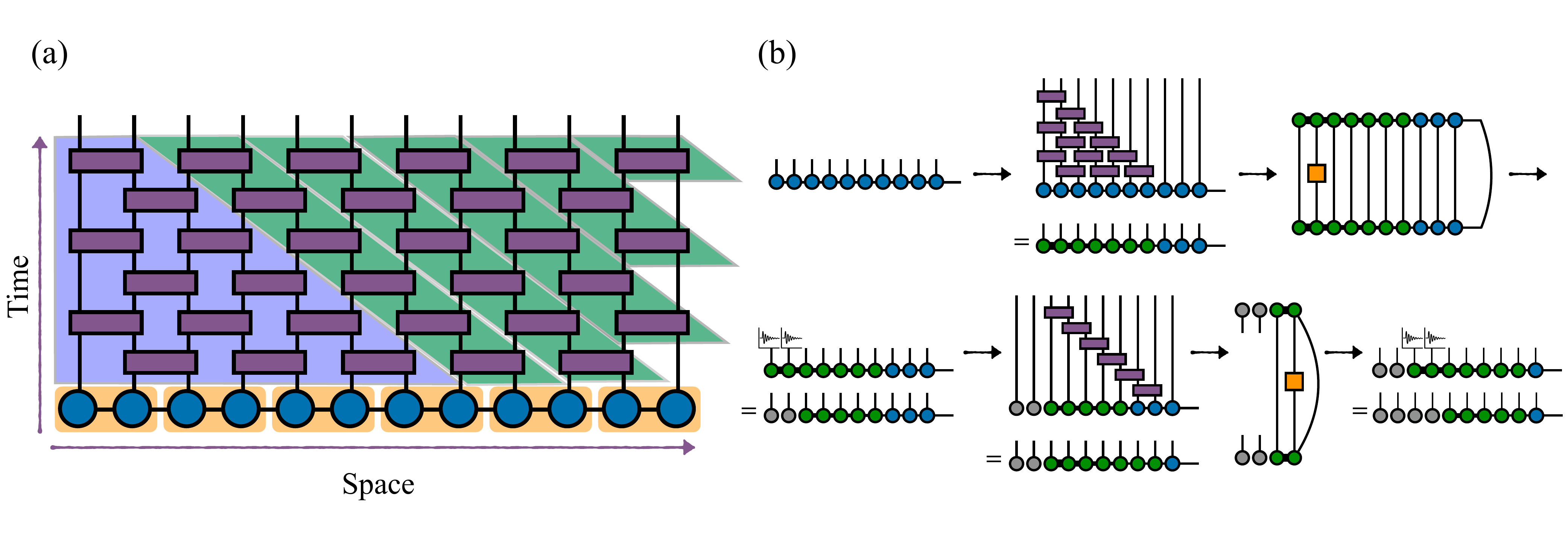} 
    \caption{Schematic tensor network representation of the SEBD framework for real-time dynamics. (a) Layout for the kicked Ising model, shown for $N = 12$ sites for visual clarity, evolved to time $t=3$ Floquet periods using a Trotter step $\Delta\tau = 1$. All numerical simulations in the main text are performed on significantly larger systems. The time evolution is decomposed into a single left light cone and a sequence of diagonal cones, each corresponding to a causal patch associated with a two-site unit cell. This decomposition preserves causal structure and is compatible with both finite and half-infinite systems; a finite chain is shown here for illustration.
    (b) Illustration of the SEBD- and holoQUADS-inspired method for time evolution and estimation of local observables used in this work, and discussed further in Section~\ref{sec:combining}.}
    \label{fig:sebd_local}
\end{figure*}

%
%
%
%
\section{Methods and Models}  \label{sec:models_methods} 

In this section, we first review state-of-the-art tensor network techniques for simulating quantum circuits using MPS. We then discuss the main approach used in this paper. Particular emphasis is placed on the ``entangled measurement'' optimization, which substantially improves sampling efficiency by reducing the number of trajectories required to achieve a given precision.

\subsection{Classical MPS Algorithms for Dynamics of One-Dimensional Systems}

In this article, we focus on one-dimensional (1D) quantum many-body systems with short-range interactions, evolving under continuous Hamiltonian dynamics or discretized Floquet dynamics. For such systems, the time-evolution operator can be factorized into a quantum circuit to a good approximation. This approximation can be implemented via the Trotter-Suzuki decomposition, which introduces controllable discretization errors that scale polynomially with the time step size~\cite{Suzuki1976, MacLachlan:1995, Berry2007, Hatano2005}.

We will represent the many-body wavefunction undergoing time evolution as an MPS tensor network. Real-time evolution is implemented by applying the Trotterized time-evolution operator as a sequence of two-site gates acting layer by layer, followed by a truncation step after each gate that compresses the MPS while preserving its structure. This procedure defines the TEBD algorithm, which is the ``gold standard'' method for simulating circuits with MPS tensor networks \cite{Vidal2004, Verstraete2004, White2004, Daley2004}. (Other methods exist for evolving MPS that do not involve circuits~\cite{Paeckel2019,Haegeman_PhysRevLett.107.070601,Lubich_2013_dynamical,Haegeman_PhysRevB.94.165116}, but these have similar computational scaling as TEBD for the models we consider and will not be the focus of this work.)

While TEBD is a controlled and accurate algorithm, its computational cost is tied to the bond dimension of the MPS. The bond dimension constrains the amount of entanglement that can be faithfully represented in an MPS state. Thus, as entanglement increases under the time evolution of generic closed systems, the bond dimension must also be increased to preserve accuracy. In many cases of interest, such as quench dynamics, the growth of bond dimension can be exponential in time. Such an exponential growth of computational cost imposes strong limitations on TEBD, restricting its applicability to short and intermediate times in the worst cases.

\subsection{Combining Gate Evolution and Mid-Circuit Measurement \label{sec:combining}}

Viewing time evolution through the lens of quantum circuits aligns classical tensor network simulations with those performed on quantum hardware. Both approaches can execute finite-depth quantum circuits and perform projective measurements of individual qubits or sites. More surprisingly, algorithmic developments originally designed for quantum hardware can often be repurposed to accelerate classical simulations~\cite{GarciaRipoll2021quantuminspired,Chen2023PRX}. The method studied in this paper is one such example.

In both classical and quantum contexts, there is the freedom to apply the circuit along diagonals aligned with the causal light-cone structure, rather than strictly in horizontal time layers---see Fig.~\ref{fig:sebd_local}(a). In such a diagonal or light-cone approach, certain qubits reach the final time before others, meaning the reduced density matrix (RDM) of these qubits becomes independent of the yet-to-be-evolved portions of the system.

Qubits at the final time can be collapsed via projective measurements which yield a faithful ``snapshot'' of the final-time state. Once measured, their local states factorize from the system. The use of mid-circuit measurements has been proposed for recycling qubit resources in quantum simulations where it is known as the holoQUADS algorithm~\cite{Foss-Feig2021, Chertkov2022} and for reducing entanglement in classical simulations of shallow two-dimensional quantum circuits, where it is referred to as the SEBD algorithm~\cite{Napp2022}. These ideas build on proposals for ``holographic'' or ``qubit-efficient'' quantum circuits~\cite{Kim2017_Robust,Kim2017_Noise,Huggins_2019} and the sequential generation of matrix product states using finite quantum resources~\cite{Schon2005}.

Motivated by these developments, we adapt the holoQUADS and SEBD algorithms to the setting of quantum Hamiltonian time evolution, resulting in a quantum-inspired space-time evolution scheme, which we will sometimes refer to simply as ``SEBD.''

One of our motivations is to observe and understand the amount of entanglement that would actually be occurring on quantum hardware when following such protocols. Another motivation is whether SEBD yields classical computational benefits for simulating quantum Hamiltonian dynamics. While such benefits are well established for shallow circuits, this question is less obvious in the continuous-time limit (i.e., small time steps), where the quantum circuit for Hamiltonian evolution can become arbitrarily deep and the light cones very broad, making the potential for classical benefits less clear. Nevertheless, we will show that SEBD does in fact provide benefits. Moreover, the SEBD framework naturally accommodates extensions and optimizations---notably the \emph{entangled measurement} approach which we describe in more detail below---that reduce the number of samples needed by orders of magnitude compared to na\"ive averaging of the post-measurement observations.

\begin{figure}
    \centering  \hspace{-0.1in} 
    \includegraphics[width=0.48\textwidth]{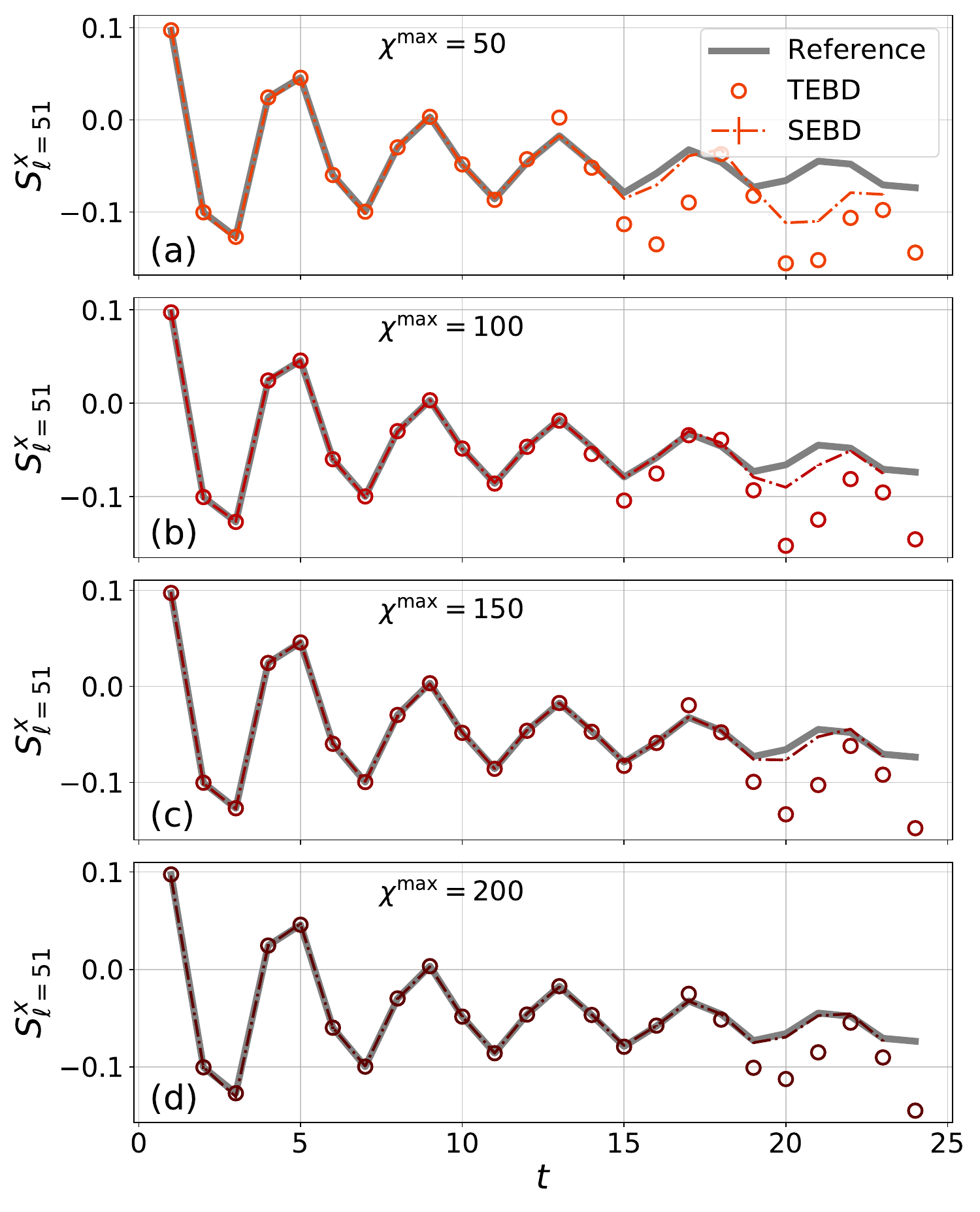}
    \caption{Benchmark of SEBD with entangled measurements against TEBD for real-time dynamics of the 1D kicked Ising model. (a)-(d) Time evolution of the local spin expectation value $\langle S^{x}_{\ell=51}(t) \rangle$, computed using SEBD (dashed lines) and TEBD (solid symbols) at increasing bond dimension cutoffs $\chi^{\mathrm{max}} = 50, 100, 150,$ and $200$. The solid blue line denotes the converged TEBD reference at $\chi^{\max} = 2000$, with convergence verified via bond-dimension extrapolation. SEBD simulations are performed on a chain of $N=100$ sites, using a truncation threshold $\epsilon=10^{-8}$ and averaging over $N_{s}=8000$ stochastic trajectories, making the statistical SEBD error bars smaller than the plotting linewidth.}
    \label{fig:Fig2}
\end{figure}

\subsection{Description of the Method}

To detail the holoQUADS- and SEBD-inspired method employed in this work, we adopt a first-order Trotter decomposition of the time-evolution operator as a proof of principle, $\mathrm{U(\Delta\tau) = U_{even}(\Delta\tau) U_{odd}(\Delta\tau) + \mathcal{O}(\Delta\tau^2)}$, as depicted in Fig.~\ref{fig:sebd_local}(a). The method generalizes straightforwardly to higher-order Trotter decompositions, provided they can be expressed in a brick-wall type circuit structure. The quantum state is represented as an MPS and evolved by sequentially applying all gates within the causal light cone of a selected two-site unit cell, also highlighted in the figure. Once the physical sites of the cell reach the target evolution time, projective measurements are performed, disentangling the measured sites from the rest of the system. This process is repeated across the chain, such that each site is evolved and measured exactly once per trajectory. After all sites have been evolved and measured, and estimators of physical observables have been collected, the algorithm restarts from the beginning to compute the next independent sample.

To illustrate the procedure, Fig.~\ref{fig:sebd_local}(b) depicts the steps of the algorithm using tensor diagram notation. The evolution begins with the left light cone---a triangular region extending from the left edge (purple shaded)---initially covering sites $\ell = 1$ to $\ell = 8$ in the schematic example shown.  Once the first two sites reach the final time, estimators of observables can be evaluated---we discuss two options for doing so below. More complex observables such as time-dependent correlators can also be measured within this framework, as discussed in Section~\ref{sec:correlation_functions}.

The algorithm proceeds by projectively measuring the current two sites to disentangle them from the rest of the system, as shown in the  lower panel of Fig.~\ref{fig:sebd_local}(b). The algorithm then advances to the next unit cell: gates within the subsequent diagonal cone are applied, followed by observable measurement and projective measurement of the sites.

The process is repeated sequentially across the lattice until all sites have been evolved to the final time and measured. The spatial footprint of the simulation cell is set by the width of the causal light cone, which scales as $\mathcal{O}(2t/\Delta\tau)$, where $t$ denotes the total evolution time and $\Delta\tau$ is the Trotter step size.

\subsection{Entangled Measurement Optimization}

Because the algorithm above samples each site at the final time, a straightforward way to estimate observables is to use the collected samples and average them over shots of the algorithm. We refer to this baseline approach as sampling and averaging ``bitstrings.'' While straightforward to implement, this approach requires a large number of samples to achieve acceptable precision (usually thousands of independent samples). In the quantum hardware setting, the inefficiency of computational-basis sampling is well known and has motivated innovative ideas such as shadow tomography~\cite{huang2020predicting,Akhtar2023scalableflexible,Bertoni2024}, which can dramatically reduce the number of samples required by performing additional circuit evolution and classical post-processing.

Here, we are interested in taking advantage of our classical simulation setting where we have a tensor network representation of the entire quantum state at each step of the SEBD algorithm, between each light-cone evolution and sampling step. A much more efficient strategy is to collect estimates of observables \emph{before} projectively measuring, as shown on the right-hand side of Fig.~\ref{fig:sebd_local}(b).

We emphasize that this strategy---referred to as \emph{entangled measurement}---can only be performed efficiently in a classical simulation context, since it involves computing the numerically exact expectation values of entangled quantum states (i.e., the intermediate states just before each projective measurement step). These expectation-value measurements can be carried out efficiently using conventional MPS techniques.

Most importantly, we find that the EM approach reduces the required number of samples to achieve a given precision by at least an order of magnitude---typically a few hundred samples suffice for precision around $10^{-3}$ compared to several thousand for estimators based on the post-projected-measurement states (na\"ive bitstring sampling approach)---as demonstrated in Fig.~\ref{fig:Fig8} and further corroborated in Appendix~\ref{appendix:RDM_sampling} and~\ref{appendix:heisenberg}. Throughout various figures below, such as Fig.~\ref{fig:sample_variance} and Fig.~\ref{fig:Fig8}, we quantitatively compare how the efficiency of EM compares to the na\"ive bitstring approach in terms of the number of samples needed (more precisely, the sample variance obtained with each strategy).

\begin{figure}
    \hspace{-0.1in}
    \includegraphics[width=0.5\textwidth]{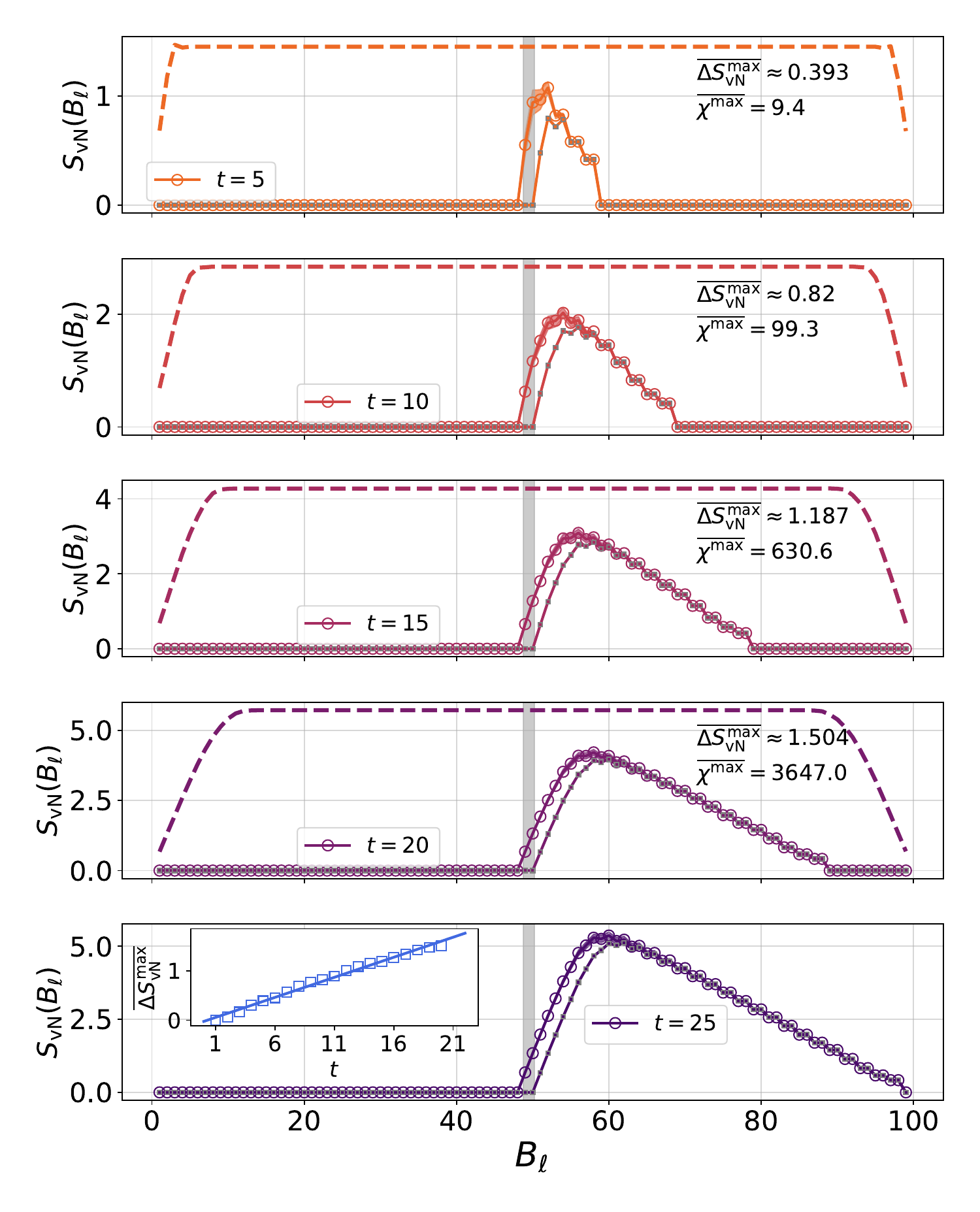}
    \caption{Suppression of entanglement growth via projective measurements in SEBD. Panels (a)-(e) show sample-averaged spatial profiles of the von Neumann entanglement entropy $S_{\mathrm{vN}}(B_{\ell})$ at representative times, comparing SEBD (before and after applying projective measurements) to TEBD. Entropy is computed across bond $B_{\ell}$, which partitions the chain between sites $\ell$ and $\ell+1$. Projective measurements are applied to sites $\ell=49$ and $\ell=50$ immediately prior to sampling in SEBD, resulting in a sharp drop in $S_{\mathrm{vN}}$ to zero at $B_{\ell}=49$ and $B_{\ell}=50$, consistent with complete local disentanglement of the measured sites. Dashed lines denote TEBD reference profiles at corresponding times. Each panel reports the sample-averaged peak entanglement gap $\overline{\Delta S^{\max}_{\mathrm{vN}}}$ and the corresponding peak bond dimension difference $\overline{\Delta\chi^{\max}}$ between SEBD and TEBD. The inset of panel (e) shows that $\overline{\Delta S^{\max}_{\mathrm{vN}}}$ grows approximately linearly with time. Since the bond dimension scales as $\chi \varpropto e^{S_{\mathrm{vN}}}$, this linear entanglement separation implies an exponential relative late-time advantage for SEBD in the required bond dimension per sample. Further analysis is provided in Appendix~\ref{appendix:ising_chi}.}
    \label{fig:sebd_entanglement}
\end{figure}

\subsection{Model Systems Investigated}

To study the above algorithm, we investigate the real-time dynamics of the kicked Ising model and the Heisenberg model. The kicked Ising model is a paradigmatic Floquet system recently realized in quantum simulation experiments~\cite{Chertkov2022}. It consists of a 1D spin chain subject to a periodically applied transverse field and a static longitudinal field $h$, which breaks integrability and drives the system into a chaotic, thermalizing regime~\cite{Bertini2018, Bertini2019}. The stroboscopic time evolution is generated by the Floquet operator constructed from the time-dependent Hamiltonian:
\begin{align} \label{eq:Eq1}
    H(t) & = \sum_{i=1}^{N-1} J \sigma_{i}^{z} \sigma_{i+1}^{z} +  \sum_{i=1}^N h\sigma_{i}^{z}  \nonumber \\
    & \mbox{} + \frac{\pi}{4} \sum_{i=1}^N \sum_{ n \in \mathbb{Z}} \delta(t - n) \sigma_{i}^{x},
\end{align}
where $\sigma_{i}^{\alpha}$ are Pauli operators acting on site $i$, and the transverse field is applied through periodic delta-function kicks at integer times. Setting the Ising coupling to \mbox{$J = \pi/4$} places the model at a self-dual point, where the dynamics are dual-unitary---unitary in both time and space directions---rendering the system maximally chaotic yet analytically tractable~\cite{Bertini2018, Bertini2019, Piroli2020}. To probe the performance of our algorithm in a more generic setting, we study it in the non-integrable, ergodic regime by setting $J = \pi/8$.

To provide a contrast to the kicked Ising model, whose symmetries allow time-evolution circuits with large step sizes and narrow light cones, we also investigate the 1D $S=1/2$ Heisenberg model, where accurate evolution requires many gates with small time step sizes, resulting in a large circuit depth and wide light cones. The Hamiltonian for this model is 
\begin{eqnarray}
    H = \sum_{i = 1}^{N - 1} \left( S_{i}^{z}S_{i+1}^{z} + \frac{1}{2} S_{i}^{+}S_{i+1}^{-} + \frac{1}{2}S_{i}^{-}S_{i+1}^{+} \right),
\end{eqnarray}
where the sum runs over all nearest-neighbor sites in a chain of length $N$. The Heisenberg model serves as a cornerstone for understanding quantum magnetism in low-dimensional systems.

\begin{figure}
    \hspace{-0.2in}
    \includegraphics[width = 0.5\textwidth]{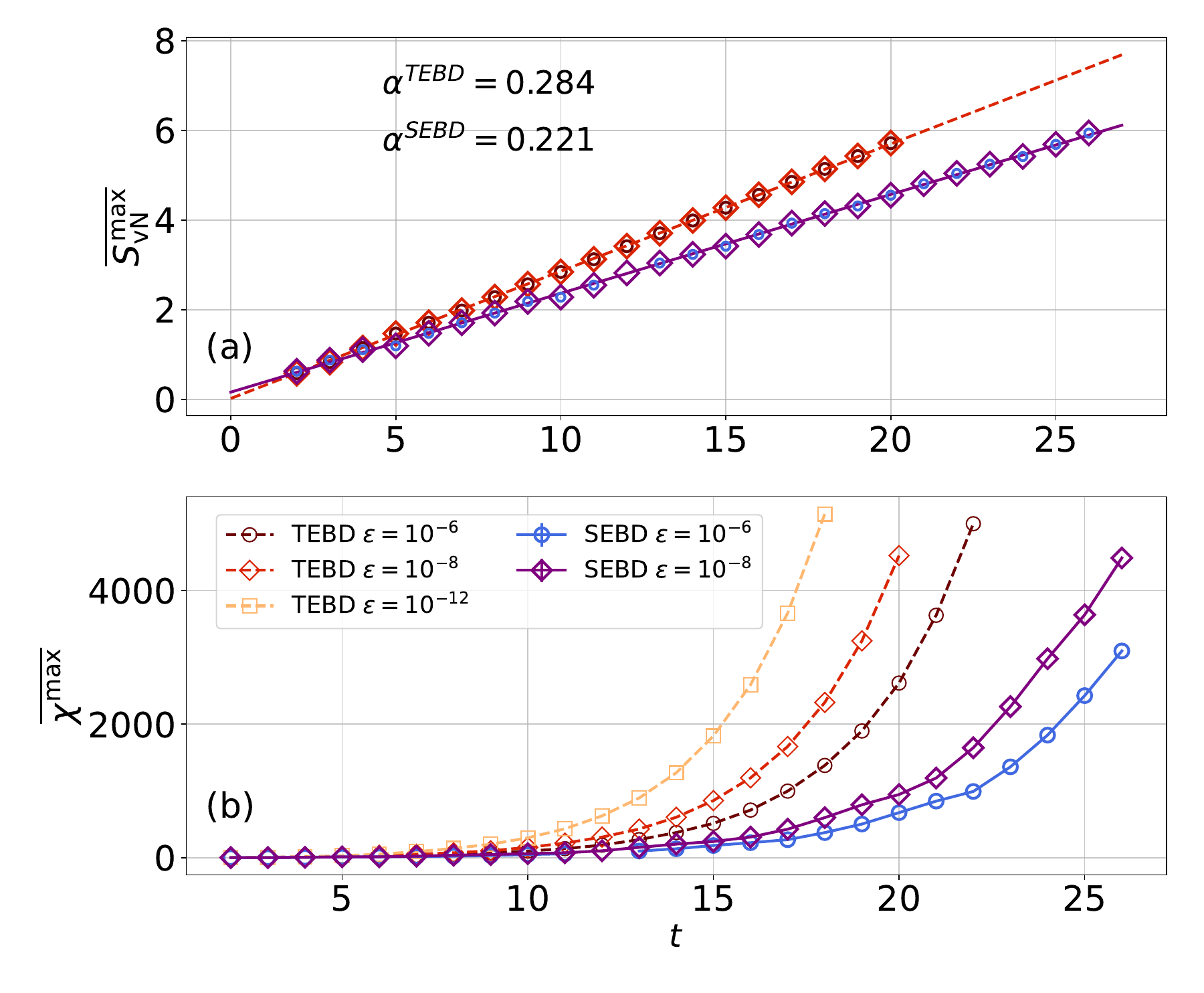}
    \caption{Time dependence of maximum entanglement and bond dimension in SEBD versus TEBD. (a) Sample-averaged maximum von Neumann entanglement entropy $\overline{S^{\max}_{\mathrm{vN}}}$ and (b) sample-averaged maximum bond dimension $\overline{\chi^{\max}}$ as a function of time, computed using SEBD and TEBD under varying truncation thresholds $\epsilon$. At fixed $t$, SEBD yields substantially lower entanglement and bond dimension across all values of $\epsilon$, even though both methods exhibit asymptotic exponential scaling of bond dimension with time. The entanglement gap $\overline{\Delta S^{\max}_{\mathrm{vN}}}$ and the corresponding bond-dimension gap $\overline{\Delta \chi^{\max}}$ grow monotonically with time, underscoring the increasing late-time advantage of SEBD in terms of computational efficiency. Simulations are performed on a kicked Ising chain of length $N = 100$, initialized in a N\'{e}el state, with integrability broken by a longitudinal field $h=0.2$.}
    \label{fig:Fig4}
\end{figure}

%
%
%
%
\section{Dynamics of Entanglement within SEBD}  \label{sec:entanglement_dynamics}

In this section, we systematically investigate the impact of projective measurements on entanglement dynamics in both the 1D kicked Ising and Heisenberg models, using the framework discussed in the previous section, which is inspired by holoQUADS and SEBD. Our results have direct implications for recent holographic quantum simulation experiments on trapped-ion processors~\cite{Chertkov2022}, in the sense that we can observe the levels of entanglement present in such experiments, assuming a closed-system approximation. We also find that the protocol generalizes well to continuous-time dynamics, as illustrated through simulations of the Heisenberg model.

\subsection{Performance of SEBD versus TEBD  \label{ssec:ising}}

Figure~\ref{fig:Fig2} compares the time evolution of the local observable $S^{x}_{\ell}(t)$ at the center of the lattice ($\ell = 51$), computed using SEBD and TEBD for various choices of maximum bond dimension limits. At fixed truncation threshold and maximum bond dimension, SEBD systematically preserves accuracy over substantially longer timescales than TEBD, indicating that TEBD saturates its bond dimension or entanglement limit sooner. For example, at $\chi^{\max} = 200$, SEBD remains in excellent agreement with the high-accuracy reference calculation throughout the entire simulation window, while TEBD with the same bond dimension limit deviates substantially once $t \gtrsim 19$. All SEBD results are averaged over $N_{s} = 8000$ samples, making sampling errors much smaller than MPS truncation errors.

Taken together, these results demonstrate that, at fixed per-sample computational cost, SEBD can access longer evolution times compared to the usual TEBD approach. This does not necessarily imply that SEBD uses fewer resources overall, since SEBD must be repeated to accumulate sufficient samples. But because the outer sampling loop of SEBD parallelizes in a trivial way, with enough computational nodes running in parallel, SEBD can reach long target times in less ``wall clock'' or actual time than TEBD running on a single node.

\begin{figure}
    \centering \hspace{-0.2in}
    \includegraphics[width=0.5\textwidth]{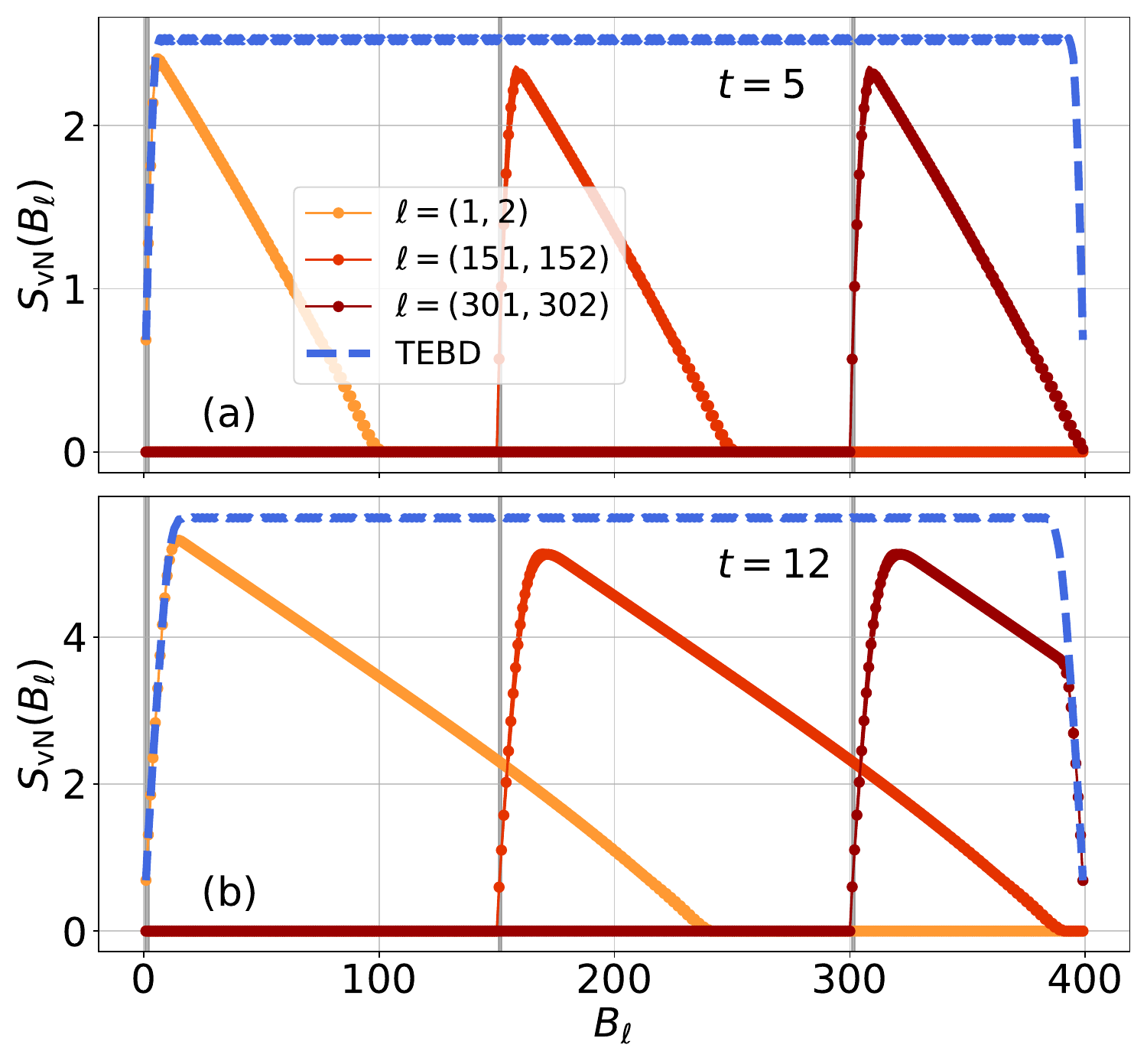}
    \caption{Entanglement suppression in the Heisenberg model via SEBD. Spatial profiles of von Neumann entanglement entropy $S_{\mathrm{vN}}(B_{\ell})$ at (a) $t=5$ and (b) $t=12$, computed using SEBD (solid circles) and TEBD (dashed lines) for the 1D spin-$1/2$ Heisenberg model. SEBD simulations proceed from left to right, following the entanglement light-cone structure. Entropies are recorded immediately before projective measurements at representative unit cells $(\ell, \ell + 1)=(1, 2), (151, 152), (301, 302)$. SEBD suppresses entanglement growth by disentangling measured sites and yields a widening gap in peak entanglement relative to TEBD over time. Simulations are performed on a chain of $N = 400$ sites, initialized in a N\'{e}el state, with $N_{s}=1000$ sample trajectories, Trotter step size $\Delta\tau = 0.1$, and truncation threshold $\epsilon = 10^{-6}$.}
    \label{fig:heisenberg_entanglement}
\end{figure}

\begin{figure}
    \hspace{-0.2in}
    \includegraphics[width=0.5\textwidth]{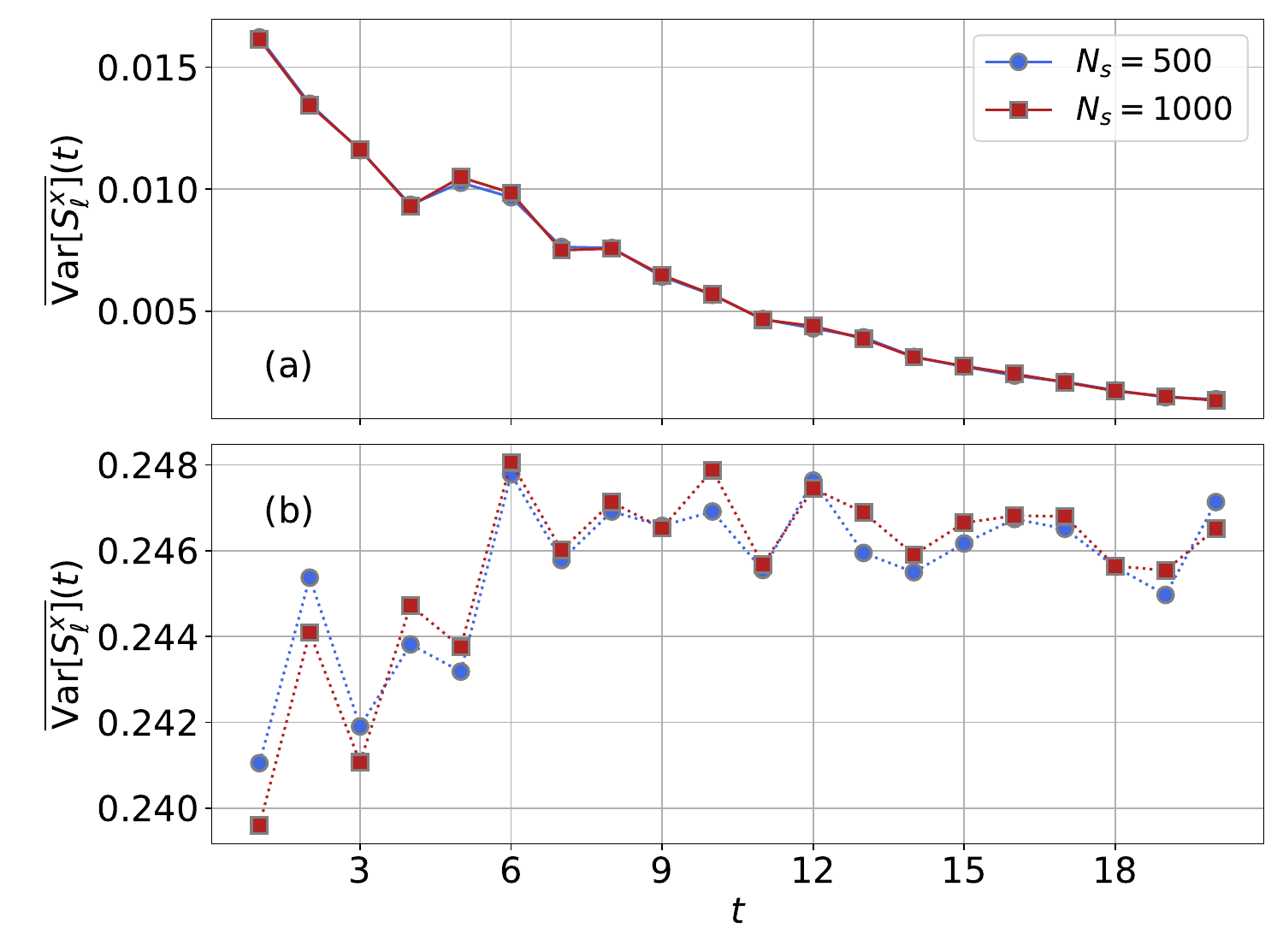}
    \caption{Time dependence of sampling variance in estimating local observables using (a) EM and (b) bitstring sampling protocols. Plotted is the spatially averaged sampling variance of $\langle S^{x}_{\ell} \rangle$ as a function of real time, computed over a spatial window of size $L = 10$ centered at $\ell=50$, for fixed sample sizes $N_{s}=500$ and $N_{s}=1000$. In panel (a), EM sampling yields a pronounced reduction in variance with increasing time. This time-dependent improvement in sampling efficiency contrasts sharply with the behavior in panel (b), where bitstring sampling exhibits a time-independent variance plateau at approximately $\sim 0.25$. These results highlight the statistical advantage of EM sampling in the presence of strong entanglement growth at late times.}
    \label{fig:sample_variance}
\end{figure}

\subsection{Time Dependence of Entanglement Entropy  \label{ssec:ising}}

To elucidate the mechanism underlying the reduced bond dimensions and enhanced performance (per sample) observed in SEBD compared to TEBD, we study the spatial profile of the entanglement entropy throughout the SEBD algorithm in Fig.~\ref{fig:sebd_entanglement}. For the systems we consider, the entanglement of the full state at each time step, and hence the TEBD entanglement (dashed curves in Fig.~\ref{fig:sebd_entanglement}), is flat throughout the bulk.

In contrast, SEBD yields a dynamically evolving, dome-shaped entanglement profile whose spatial extent scales as $\mathcal{O}(2t/\Delta\tau)$. This shape is a direct consequence of SEBD's light-cone-based evolution: two-site unit cells are time-evolved sequentially from left to right, and projective measurements are performed immediately upon each cell reaching the target time. These projective measurements, equivalent to mid-circuit measurements in quantum simulations, disentangle the measured sites from the rest of the system, quenching entanglement and preventing its accumulation. As a result, entanglement remains localized within a light cone that propagates across the system.

Fig.~\ref{fig:sebd_entanglement} quantitatively illustrates the effect of projective measurements on entanglement entropy by comparing the von Neumann entanglement entropy $S_{\mathrm{vN}}(B_{\ell})$ before and after projections at sites $\ell = 49$ and $\ell = 50$. Entanglement entropy is evaluated across each bond $B_{\ell}$, which bipartitions the chain between sites $\ell$ and $\ell + 1$. Immediately after the measurement, $S_{\mathrm{vN}}(B_{\ell})$ across bonds $B_{49}$ and $B_{50}$ collapses to zero, as shown by open circles and solid squares, reflecting the local disentanglement induced by projection.

Most importantly, SEBD consistently yields a significantly lower peak entropy than TEBD and this entropy difference, $\overline{\Delta S_{\mathrm{vN}}^{\max}}$, grows roughly linearly with evolution time, as shown in the inset of Fig.~\ref{fig:sebd_entanglement}(e). Thus the entanglement entropy suppression advantage of SEBD becomes even more significant at later physical times. The entanglement reduction directly translates into reduced bond dimensions and improved computational efficiency. It is also important to note that while the regions of non-zero entanglement have a very broad shape, especially at longer times and in the continuous-time limit, because computational costs scale as $\mathcal{O}(\chi^3)$, it is the height of the entanglement peak, rather than its width, that primarily determines the cost of a calculation.

To elucidate the relationship between the (sample-averaged) maximum entanglement entropy, $\overline{S^{\max}_{\mathrm{vN}}}$, and the corresponding maximum bond dimension, $\overline{\chi^{\max}}$, we analyze their time dependence in Fig.~\ref{fig:Fig4} (a)-(b). Panel~(a) shows $\overline{S^{\max}_{\mathrm{vN}}}$ as a function of evolution time for both SEBD and TEBD, across multiple truncation thresholds $\epsilon$. In both methods, peak entanglement entropy growth is approximately linear in time and insensitive to the $\epsilon$ used. Linear fits yield growth rates $\alpha^{\mathrm{TEBD}} = 0.284$ and $\alpha^{\mathrm{SEBD}} = 0.221$, confirming that SEBD systematically suppresses entanglement relative to TEBD. This reduced growth rate delays the onset of rapid bond dimension growth. For instance, at $\epsilon = 10^{-8}$, TEBD reaches $\chi^{\max} \sim 4000$ by $t=20$, whereas SEBD requires an extended evolution time $t \approx 26$ to reach a comparable bond dimension. This time delay persists across looser truncation thresholds (e.g. $\epsilon = 10^{-6}$), highlighting the robustness of SEBD in controlling entanglement and extending the practical simulation window at fixed computational cost.

\subsection{Simulating Continuous Time Evolution with SEBD  \label{ssec:heisenberg}}

A non-trivial case of the SEBD algorithm arises in the limit of deep circuits. For a generic deep circuit, entanglement can proliferate across the entire system, too late for the projective measurement of any sites to have a beneficial effect. The case of continuous-time Hamiltonian evolution leads to deep circuits when small time steps are used, raising similar concerns. However, as we demonstrate below, SEBD retains its advantage even in this deep-circuit regime.

\begin{figure*}
    \centering
    \includegraphics[width=1\textwidth]{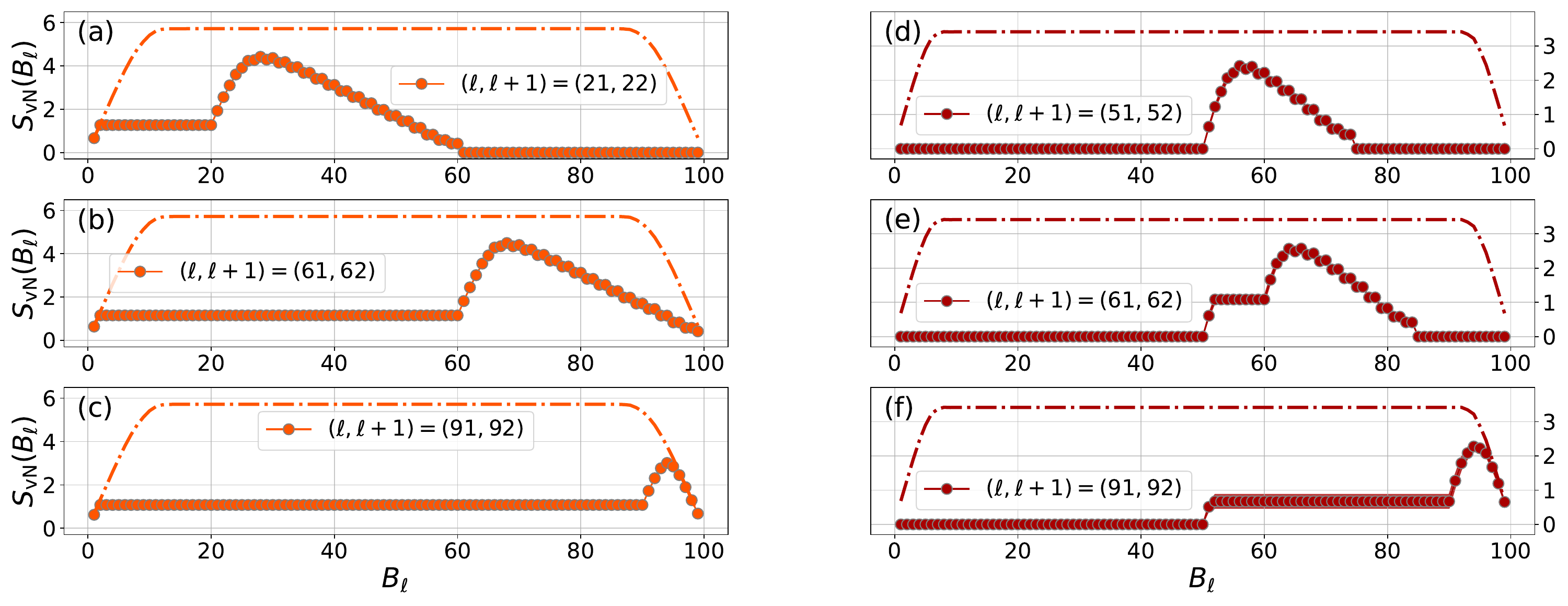}
    \caption{Light-cone dynamics of the entanglement entropy dome during measurement of equal-time correlators in SEBD. Spatial profiles of the von Neumann entropy $S_{\mathrm{vN}}(B_{\ell})$ evaluated within the SEBD framework during the sequential measurement of equal-time spin-spin correlation functions. Panels (a)-(c) in the left column show $C^{zz}(\ell = 1, \ell')$ at $t = 20$, using the left boundary site $\ell = 1$ as the fixed reference. Panels (d)-(f) in the right column display $C^{xx}(\ell = 51, \ell')$ at $t = 12$, with the central site $\ell = 51$ as the reference. In both cases, $S_{\mathrm{vN}}(B_{\ell})$ is computed across bond $B_{\ell}$, which bipartitions the system into two subsystems. As measurements proceed unit cell by unit cell, the entropy dome propagates along the physical sites and tracks the moving light cone. Shaded regions indicate one standard deviation over $1000$ sampled trajectories. Simulations are performed on a chain of $N = 100$ sites, with a longitudinal field $h=0.2$ breaking integrability.}
    \label{fig:Fig7}
\end{figure*}

To investigate the continuous-time limit, we consider the spin-$1/2$ Heisenberg chain using a refined time step size of $\Delta\tau = 0.1$. The smaller time step broadens the spatial extent of each light cone and increases the circuit depth, thereby enlarging the simulation unit cell by approximately an order of magnitude compared to the kicked Ising case.

Figure~\ref{fig:heisenberg_entanglement} shows the spatial propagation of the entanglement dome at representative times $t=5$ and $t=11$ during the evolution and measurement of unit cells containing sites $\ell=(1, 2)$, $\ell=(151, 152)$, and $\ell=(301, 302)$ in the Heisenberg model. As anticipated, the width of the dome scales as $\mathcal{O}(2t/\Delta\tau)$, substantially larger than that in simulating the kicked Ising model. Indeed, as seen in the lower panel of Fig.~\ref{fig:heisenberg_entanglement} at $t=11$, the entanglement dome spreads over roughly half the system.

Yet we find that the entanglement remains significantly lower than the baseline of TEBD throughout the time evolution. The need to use a deeper circuit does not substantially increase the typical entanglement: it only creates a long ``tail'' of low entanglement that contributes insignificantly to the cost for the classical simulation. This is because, while the circuit becomes deeper, the gates become closer to the identity and less entangling individually. Thus, in the light cone tails, the entanglement remains low.

Moreover, as in the kicked Ising case, the entanglement entropy gap---defined as the difference in peak von Neumann entropy between SEBD and TEBD---increases with time, mirroring the trend observed in the kicked Ising model. This growing entropy gap demonstrates that SEBD’s entanglement-suppressing mechanism is not limited to Floquet Hamiltonians, and we expect it to apply to generic 1D quantum systems. Because entanglement directly controls the computational cost of tensor network simulations, SEBD's ability to limit entanglement growth substantially extends accessible simulation times at fixed truncation error thresholds. We show further in Appendix~\ref{appendix:heisenberg} that these findings for the continuous-time case also translate into significantly lower bond dimensions, and thus much greater computational efficiency per sample compared to TEBD.

\subsection{Sampling Variance and Efficiency of Local Observables}

To quantitatively assess sampling efficiency, we examine the temporal evolution of the sampling variance associated with estimating local observables. As shown in Fig.~\ref{fig:sample_variance}, we focus on two representative sample sizes, $N_{s}=500$ and $N_{s}=1000$, chosen to mirror typical snapshot counts in contemporary quantum hardware experiments. We validate convergence by systematically increasing $N_{s}$ from $20$ to $1000$, observing that the time-dependent variance saturates for $N_{s} \geq 500$. As a benchmark observable, we consider the one-point function $S^{x}_{\ell}(t)$ and evaluate the spatially averaged sample variance over a central window of $L=10$ sites centered at a reference position $\ell^{r}=50$:
\begin{eqnarray}
    \overline{\mathrm{Var}[S^{x}_{\ell}]}(t) = \frac{1}{L + 1} \sum_{\ell = \ell^{r} - L/2}^{\ell^{r} + L/2} \mathrm{Var}[S^{x}_{\ell}](t).
\end{eqnarray}

This quantity is evaluated for both EM protocol and conventional bitstring sampling. For $N_{s}=1000$, both methods achieve convergence, enabling a direct comparison of sampling efficiency. Bitstring sampling exhibits a nearly time-independent variance that saturates around $\sim 0.25$, indicating the statistical uncertainty is high even at late times. In stark contrast, the variance associated with the EM protocol decreases systematically over time, falling from $\sim 0.015$ at $t=1$ to $\sim 0.002$ at $t=20$. This monotonic reduction demonstrates that, for a fixed number of samples, the EM approach gains precision as the system evolves. These findings are consistent with those shown in Appendix~\ref{appendix:RDM_sampling}, where EM achieves comparable or superior accuracy using an order of magnitude fewer samples than required by bitstring sampling.

An alternative approach we explored for reducing sampling variance was to dynamically rotate the sampling basis into the eigenbasis of the RDM at each site, just before collapsing it. While quite interesting, this approach ultimately did not yield significantly greater variance reduction than the simpler EM approach in a fixed basis. However, future work may find that RDM-basis sampling offers strong benefits in cases we did not explore. We present and discuss our findings on the RDM sampling strategy in Appendix~\ref{appendix:RDM_sampling}.

%
%
%
%
\section{Computing Correlation Functions}  \label{sec:correlation_functions}

Beyond efficiently computing one-point functions, the SEBD framework naturally extends to the evaluation of correlation functions, including both equal-time and time-dependent correlators. These measurements can be performed within the EM protocol, enabling reductions in sampling variance. In this section, we discuss strategies to obtain these correlators, demonstrate that they can be computed accurately, and explore the entanglement of the intermediate states which exhibits interesting plateaus or strings of entanglement.

\subsection{Equal-time correlators}

Computing equal-time correlation functions is a fundamental task in characterizing quantum systems undergoing Hamiltonian time evolution. Within the SEBD framework, such correlators can be evaluated efficiently, as we briefly discuss here, with additional details provided in Appendix~\ref{appendix:equal-time_correlators}.

Consider equal-time spin-spin correlators $C^{zz}(\ell, \ell')$ with a fixed reference site $\ell$. First, the reference site $\ell$ is evolved to the target time by applying all gates within its causal light cone. As subsequent gates are applied to time-evolve other sites, however, the reference site is never projectively measured and is left entangled. The remaining sites $\ell' > \ell$ are then evolved and measured sequentially, following the standard SEBD procedure. Our approach, which again involves expected values within entangled states, allows for much more efficient estimation of two-point correlators compared to na\"ive bitstring averaging of the projective measurement outcomes.

\begin{figure}
    \hspace{-0.2in}
    \includegraphics[width=0.5\textwidth]{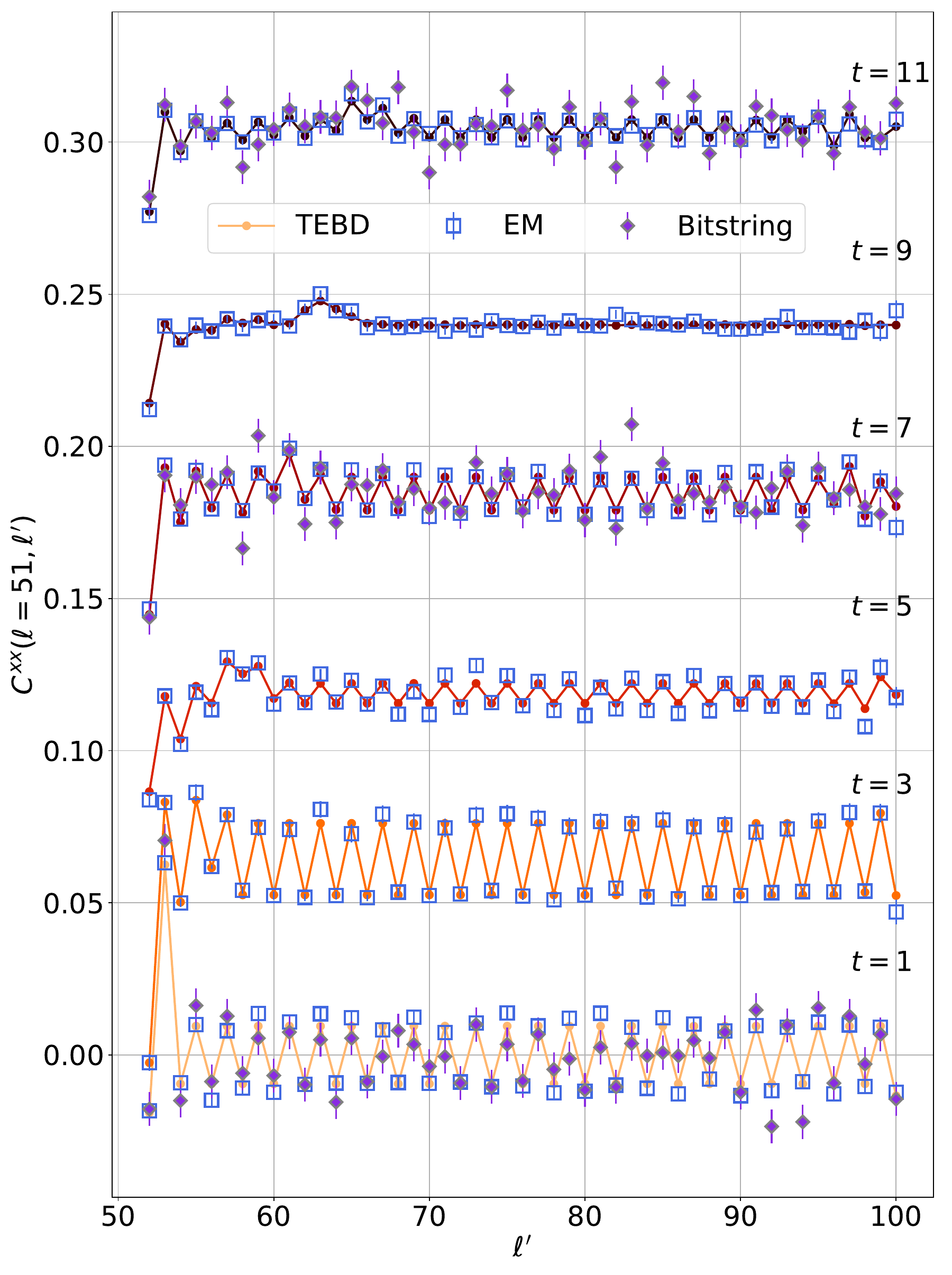}
    \caption{Sampling efficiency comparison between EM and bitstring sampling for equal-time spin-spin correlations in the kicked Ising model. Equal-time correlations $C^{xx}(\ell = 51, \ell^{\prime})$ are computed using SEBD on a chain of $N=100$ sites initialized in a N\'{e}el state. The reference site is fixed at $\ell=51$, and correlations are evaluated for all $\ell^{\prime} > \ell$. Results are averaged over $N_{s}=200$ trajectories for EM and $N_{s} = 2000$ for bitstring sampling, illustrating the sampling advantage of EM. The on-site value $C^{xx}(51, 51)$ is omitted for clarity. Traces corresponding to different evolutions times are vertically offset to facilitate comparison. EM consistently yields significantly higher accuracy with an order-of-magnitude fewer samples, underscoring its superior sample efficiency in estimating two-point correlators.}
    \label{fig:Fig8}
\end{figure}

\begin{figure*}
    \includegraphics[width=0.85\textwidth]{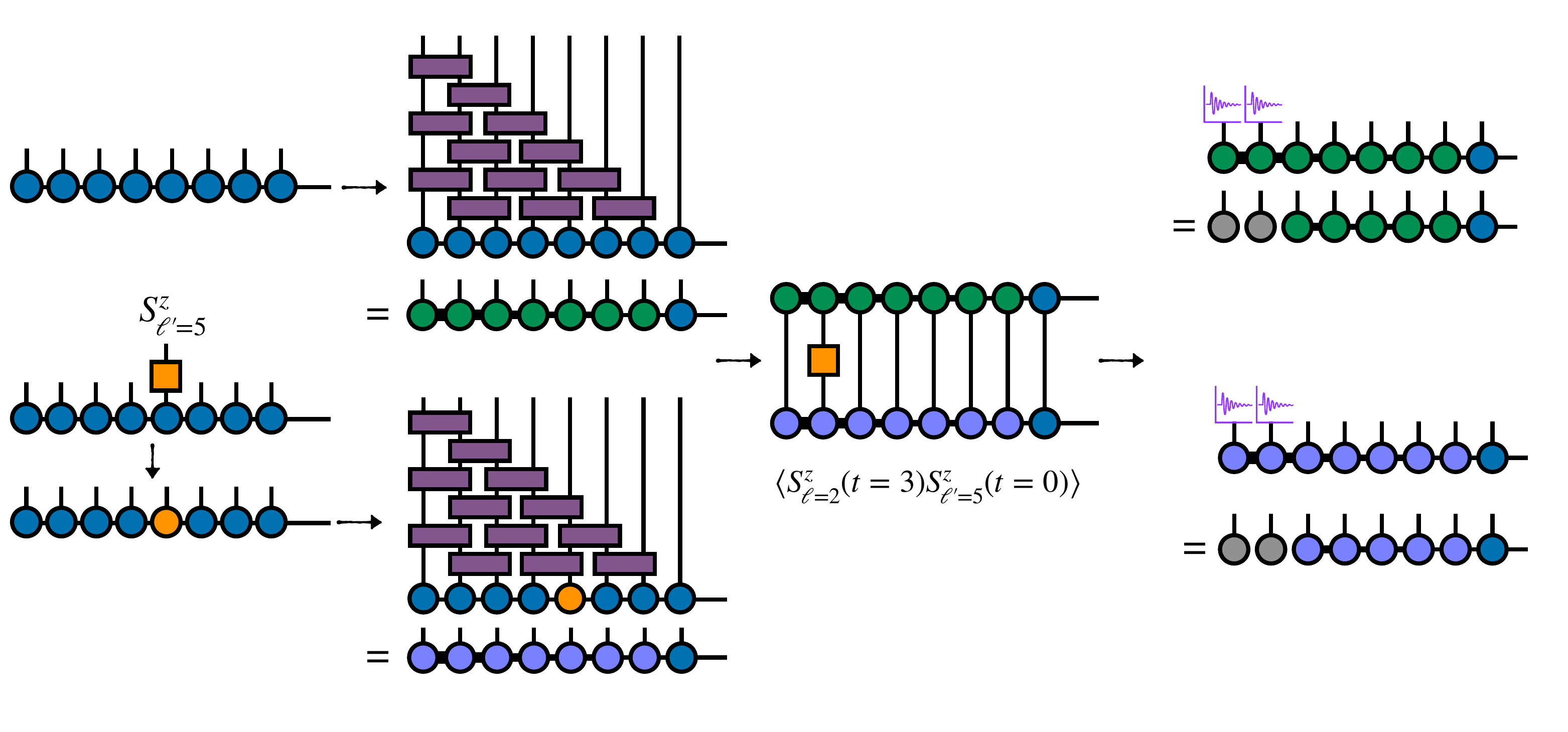}
    \caption{Tensor-network schematic of the SEBD protocol for computing unequal-time spin-spin correlation functions, e.g., $\langle S^{z}_{\ell=2}(t=3) S^{z}_{\ell'}(t=0) \rangle$. Two identical copies of the initial MPS, $|\Psi_{0}\rangle$, are independently time evolved using the SEBD algorithm. A local perturbation $S^{z}_{\ell'}$ is applied to one copy at the reference site $\ell' = 5$, yielding the modified state $(|\Psi'_{0}\rangle = S^{z}_{\ell'=5}|\Psi_{0}\rangle)$ before evolving in real-time axis. Both wavefunctions are then evolved forward in time by sequentially applying gates according to the light-cone structure. At each step, unequal-time correlators $\langle S^{z}_{\ell}(t) S^{z}_{\ell'=5}(0) \rangle$ are evaluated by computing overlaps of the form $\langle \Psi(t) | S^{z}_{\ell} | \Psi'(0) \rangle$, starting with the earliest sites $(\ell = 1, 2)$. Following each overlap evaluation, both wavefunctions are collapsed via a correlated projective measurement in a specific basis. This procedure is repeated iteratively across successive diagonals of the light cone, advancing one unit cell at a time, until the full correlator profile is reconstructed. The protocol is fully general and permits arbitrary placement of the reference site $\ell^{\prime}$, allowing computation of $\langle S^{\alpha}_{\ell}(t) S^{\alpha}_{\ell^{\prime}}(0) \rangle$ for any pairs of sites and any spin components $\alpha$.}
    \label{fig:Fig9}
\end{figure*}

To evaluate dynamics and suppression of entanglement during the computation of two-point correlators within SEBD, we examine the spatial profile of the entanglement entropy associated with evaluating $C^{zz}(\ell = 1, \ell')$ at $t = 20$. The left column of Fig.~\ref{fig:Fig7} contrasts SEBD and TEBD entanglement profiles. In SEBD, the reference unit cell ($\ell = 1, 2$) is deliberately excluded from projective measurements~\footnote{In principle, site $\ell=2$ may also be projectively measured if only the correlator $C^{\alpha\alpha}(1, \ell')$ with $\alpha \in \{x, y, z\}$ is of interest. However, since the computational benefit from disentangling a single additional site is marginal, we exclude the entire unit cell containing the reference site from projective measurement.}, resulting in a persistent entanglement plateau anchored at the left edge, with a magnitude approximately one-third of the maximal entropy in the system. As subsequent sites ($\ell' > 2$) are evolved and measured sequentially, the entanglement dome propagates rightward, tracing the causal structure of the SEBD light cone. Importantly, the peak entropy in SEBD remains substantially lower than in TEBD throughout the evolution, consistent with the suppression observed during one-point function evaluations in Fig.~\ref{fig:sebd_entanglement}(d).

The EM protocol for equal-time two-point correlation functions is fully general and accommodates arbitrary choices of the reference site. To demonstrate this flexibility, we compute $C^{xx}(\ell, \ell' \geq \ell)$ with site $\ell = 51$ fixed as the reference, as shown in the right column of Fig.~\ref{fig:Fig7}. Details of this procedure are provided in Appendix~\ref{appendix:equal-time_correlators}. As in the previous case, a finite entanglement plateau persists between the reference site and the active measurement front, encoding nontrivial correlations. As the spatial separation $|\ell - \ell'|$ increases, it is interesting to see that the height of the entanglement plateau decreases, consistent with the expected decay of correlations governed by the intrinsic correlation length $\xi$ of the system. (We conjecture that the rate of decrease of this plateau may even harbor physical information such as the correlation length.) Overall, our correlator algorithm and results confirm that SEBD with EM sampling efficiently captures the equal-time correlations while still offering entanglement suppression throughout the evolution.

To quantitatively assess the sampling efficiency and accuracy of the EM protocol for two-point correlators, we evaluate the spin-spin correlation function $C^{xx}(\ell = 51, \ell')$ and compare the results against TEBD reference values, as shown in Fig.~\ref{fig:Fig8}. Remarkably, even with a modest sample size of $N_{s} = 200$, the EM sampling reproduces the reference values accurately across all simulated times $t = 1$ to $t = 11$ (in steps of $\Delta t = 2$), with deviations remaining well within one standard error. The effectiveness of the EM approach becomes especially evident when contrasted with bitstring sampling, for example, at $t = 1$: EM faithfully captures the spatial oscillation of $C^{xx}(\ell, \ell')$, closely matching the TEBD reference across nearly all $\ell'$. In contrast, bitstring sampling with $N_{s} = 2000$---an order of magnitude more samples---produces much noisier results, with substantial error bars and systematic deviations from the reference across approximately half the sites.

\begin{figure}
    \hspace{-0.1in}
    \includegraphics[width=1\linewidth]{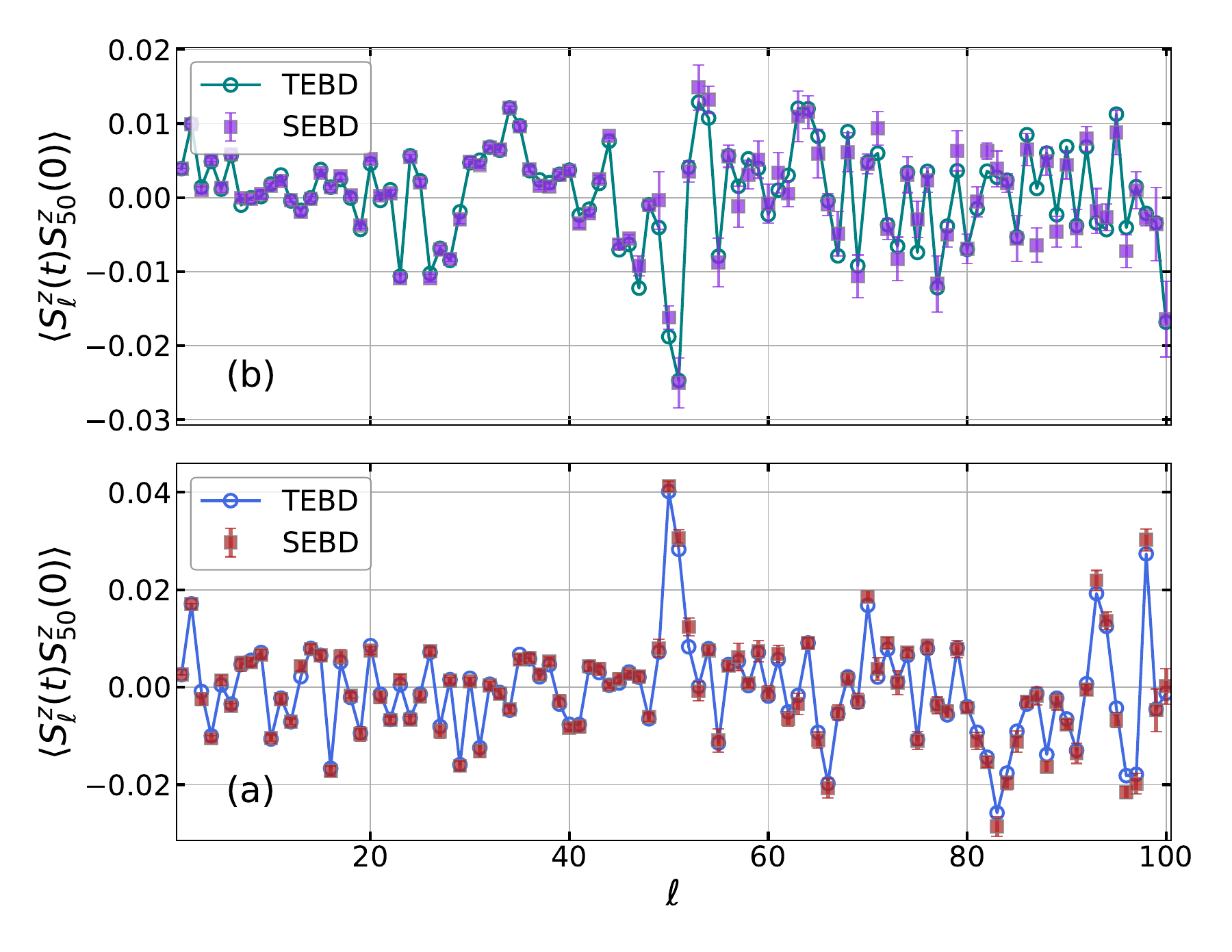}
    \caption{Benchmark of unequal-time spin-spin correlation functions $\langle S^{z}_{\ell}(t) S^{z}_{\ell^{\prime}}(0) \rangle$ computed using SEBD and TEBD. Panels (a) and (b) display the spatial profiles of unequal-time correlators at times $t=5$ and $t=6$, respectively, with the reference site fixed at $\ell^{\prime} = 50$ on a chain of length $N = 100$. SEBD results (solid squares) are obtained by averaging over $N_{s}=2000$ samples, and compared against numerically exact TEBD results (open circles). The excellent agreement across all $\ell$ demonstrates the accuracy of the SEBD protocol even at modest sample sizes. To ensure generality, both simulations are initialized from random MPS. This benchmark highlights the robustness and broad applicability of the SEBD framework for computing dynamical correlators.}
    \label{fig:Fig10}
\end{figure}


\subsection{Time-Dependent Correlation Functions}  \label{ssec:ising_unequal-time}

Beyond equal-time correlation functions, the SEBD framework naturally generalizes to the evaluation of unequal-time two-point functions of the form $\langle S^{\alpha}_{\ell}(t) S_{\ell'}^{\alpha}(0) \rangle$. Figure~\ref{fig:Fig9} illustrates the protocol for computing $\langle S^{z}_{\ell}(t) S^{z}_{\ell'}(0) \rangle$, with the reference site fixed at $\ell' = 5$. The procedure begins by initializing two identical copies of the many-body wavefunction in the state $|\Psi_{0}\rangle$. A localized perturbation is introduced at time $t=0$ by applying the spin operator $S^{z}_{\ell'}$ to one copy, resulting in the perturbed state $|\Psi'_{0}\rangle = S^{z}_{\ell'}|\Psi_{0}\rangle$.

The two states, $|\Psi_{0}\rangle$ and $|\Psi'_{0}\rangle$, are independently time-evolved under the SEBD framework, with their sampling trajectories correlated. Specifically, both wavefunctions are propagated through identical sequences of gates within the causal light cone, beginning from the leftmost unit cell. Once the sites $\ell = 1, 2$ are evolved to the target physical time $t$, the unequal-time spin-spin correlators are evaluated as
\begin{eqnarray}
    \begin{aligned}
        \langle S^{z}_{\ell}(t) S^{z}_{\ell'}(0) \rangle &= \langle \Psi_{0}| e^{iHt} S^{z}_{\ell} e^{-iHt} S^{z}_{\ell'} |\Psi_{0}\rangle \\
        &= \langle \Psi_{0}| e^{iHt} S^{z}_{\ell} e^{-iHt} |\Psi'_{0} \rangle \\
        &= \langle \Psi(t)|S^{z}_{\ell}|\Psi'(t)\rangle,
    \end{aligned}
\end{eqnarray}
for sites $\ell = 1, 2$. This expectation value is evaluated using standard MPS-MPO contraction. Following the observable evaluation, projective measurements are applied to the evolved sites in one of the wavefunctions, and the corresponding sites in the second wavefunction are collapsed onto the same measurement outcomes. This initiates the branching structure of the simulation, requiring averaging over measurement trajectories to obtain statistical estimates of the correlator.

This procedure proceeds iteratively: at each step, the appropriate diagonal light-cone gates are applied to evolve the next two-site unit cell to the target time. The unequal-time correlators $\langle S^{z}_{\ell}(t)S^{z}_{\ell'}(0) \rangle$ are then evaluated for the updated sites, followed by correlated projective measurements on both wavefunctions. Repeating this contraction-projection cycle across the chain yields the complete spatial profile of the unequal-time correlator for a fixed reference site $\ell'$, within a single sampling trajectory of the algorithm.

In Fig.~\ref{fig:Fig10}, we benchmark the accuracy of this approach in computing the unequal-time spin-spin correlation function $\langle S^{z}_{\ell}(t) S^{z}_{\ell'}(0) \rangle$, using the central site $\ell'=50$ as the reference. To demonstrate the generality of the approach, we initialize the system in distinct generic entangled states, each represented by a random MPS with bond dimension $\chi=20$, and simulate real-time evolution up to $t=5$ [Fig.~\ref{fig:Fig10}(a)] and $t=6$ [Fig.~\ref{fig:Fig10}(b)]. In both cases, SEBD results closely match TEBD reference data across all sites, with deviations remaining well within statistical error. Since the SEBD evolution proceeds from left to right, statistical noise is lowest near the left boundary and gradually increases toward the right. Nevertheless, high accuracy is maintained throughout the chain. If increased precision is required with a fixed number of samples, one can run two SEBD simulations in parallel, evolving from left to right and from right to left, respectively, and stitch together the results from each half to suppress the sampling error. Importantly, the protocol is fully general: by inserting the appropriate spin operators during the initial perturbation and final MPS-MPO contraction, the algorithm can compute arbitrary unequal-time correlators $\langle S^{\alpha}_{\ell}(t) S^{\beta}_{\ell'}(0) \rangle$ for $\alpha, \beta \in \{x, y, z\}$.

\section{Summary and Outlook} \label{sec:summary}

In this work, we develop and benchmark a measurement-assisted tensor network framework that leverages the causal light-cone structure inherent in certain quantum circuits to enable more efficient simulations of 1D quantum dynamics. Interleaving projective measurements with unitary evolution---disentangling local degrees of freedom as they reach their target physical time---systematically suppresses the growth of entanglement entropy. This, in turn, curbs the exponential increase in the MPS bond dimension. We demonstrate the efficacy and versatility of this approach in both discrete-time and continuous-time settings by applying it to Floquet dynamics in the kicked Ising model and Hamiltonian evolution in the spin-$1/2$ Heisenberg chain.

Compared to conventional MPS time-evolution methods such as TEBD, the SEBD framework reaches substantially longer real-time evolution at a fixed computational cost per sample, owing to its systematic suppression of entanglement. While SEBD introduces a sampling overhead due to its stochastic nature, this sampling overhead is trivially parallelizable, allowing one to take advantage of the slower entanglement growth in a scalable way on high-performance computing platforms with minimal effort. Moreover, the total sampling cost can be drastically reduced by employing entangled measurement, which offers substantial improvements in sampling variance.

We also demonstrated that relevant physical observables---including one-point functions, equal-time spin-spin correlations, and unequal-time correlators---can be efficiently estimated within the SEBD framework. This approach applies to both finite and half-infinite geometries, and its intrinsic entanglement suppression offers a classical benchmark for assessing whether similar protocols on quantum hardware operate within the simulable regime of tensor network methods.

These findings have implications for digital quantum simulation experiments that employ mid-circuit measurement protocols. Because direct measurement of entanglement entropy is experimentally challenging, classical simulations based on SEBD offer direct access to entanglement dynamics---providing predictive benchmarks and interpretive tools for evaluating mid-circuit measurement schemes on quantum devices and assessing whether they are reaching classically inaccessible regimes~\cite{Chertkov2022}.

Our proposal to interleave projective measurements with coherent unitary evolution to reach longer times has clear conceptual connections to other recent ideas aimed to extending the reach of classical simulations of quantum dynamics. Other methods such as density matrix truncation (DMT)~\cite{White_DMT}, dissipation-assisted operator evolution (DAOE)~\cite{Rakovszky2022}, and complex time evolution~\cite{Cao2024,Grundner2024,Cha2025} embrace the idea of replacing a faithful evolution of an entire pure state with one or more lossy evolutions that have dissipative dynamics and recover only lower-order expected values and correlation functions. The similarity of these methods and ours suggests deeper connections. The simplicity of interspersing projective measurements and sampling as in our study also points to natural extensions to other families classical methods such as neural quantum states~\cite{Medvidovic2024} or Pauli or Majorana propagation as the time-evolution method~\cite{Nys2025}.

Our study highlights the growing synergy between classical tensor network methods and digital quantum simulations---a cross-pollination that continues to drive powerful algorithmic advances. A particularly compelling direction is the integration of advanced sampling protocols into circuit cutting techniques~\cite{Peng2020, Yang2024, Harrow2025}, which could potentially suppress variance growth and enhance scalability and modularity. Extending the SEBD framework with improved sampling strategies to higher-dimensional systems, and broader classes of Hamiltonians holds promise for probing quantum phase transitions, non-equilibrium transport, and measurement-induced phase transitions. Together, these avenues underscore the potential of combining tensor network algorithms with emerging quantum simulations strategies to expand the frontiers of simulating real-time quantum many-body dynamics beyond conventional limits.

\acknowledgements
We thank Eugene~Dumitrescu for a careful reading of the manuscript and Amit~Gangapuram and Steven~White for insightful discussions. The Flatiron Institute is a division of the Simons Foundation. Part of this work was supported by the U.S. Department of Energy, Office of Science, National Quantum Information Science Research Centers, Quantum Science Center.

\clearpage
\appendix

\onecolumngrid
\begin{figure*}
    \centering 
    \includegraphics[width=1\textwidth]{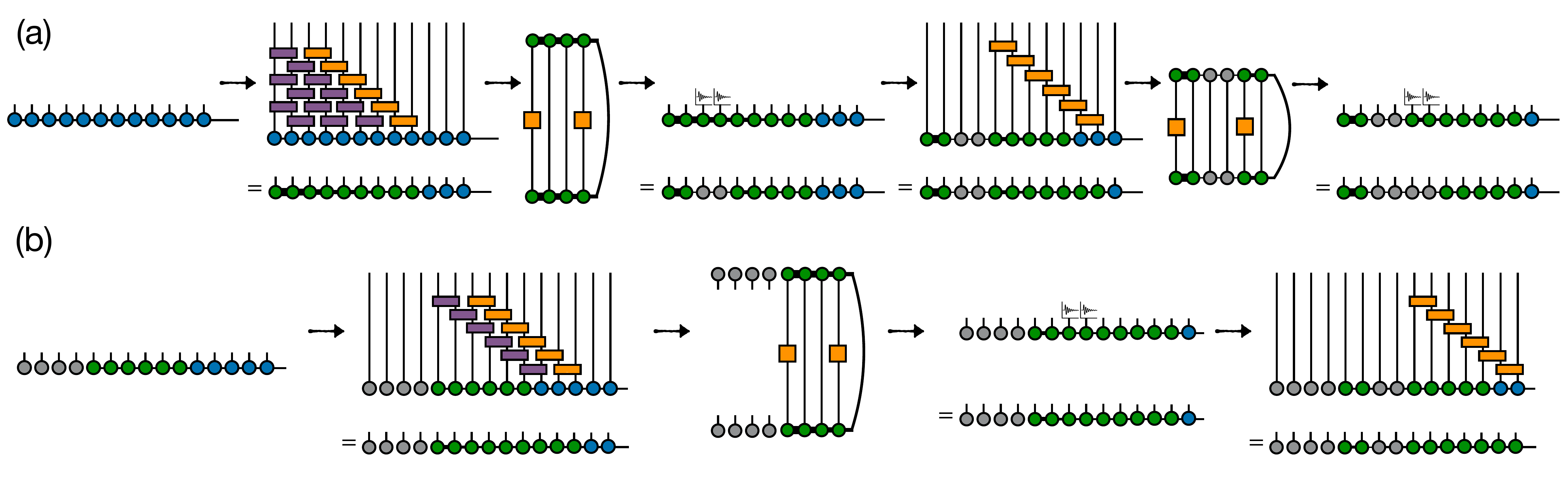} 
    \caption{Tensor network schematics for computing equal-time spin-spin correlation functions $C^{\alpha, \alpha}(\ell, \ell')$ ($\alpha = x, y, z$) within the SEBD framework. Panels (a) and (b) illustrate the entangled measurement protocol with reference sites fixed at $\ell=1$ and $\ell=5$, respectively. The system evolves under the kicked Ising model, with a transverse field applied at integer times, and is simulated for $t=3$ Floquet periods using a Trotter step $\Delta\tau=1$. Gates residing within the causal light cone of the reference site---responsible for its time evolution---are highlighted in purple.}
    \label{fig:Appendix_Fig1}
\end{figure*}
\twocolumngrid

\section{Entangled-Measurement Protocol for Equal-Time Spin-Spin Correlation Functions} \label{appendix:equal-time_correlators}

Figure~\ref{fig:Appendix_Fig1} illustrates the SEBD protocol for computing equal-time spin-spin correlation functions $C^{\alpha\alpha}(\ell, \ell')$, using two representative choices of reference site: (a) $\ell=1$ and (b) $\ell=5$. When the reference site is chosen as $\ell=1$, the initial unit cell comprising sites $\ell'=1, 2$ is evolved to the target physical time by applying all gates within its causal light cone. Crucially, this unit cell is not subject to projective measurement. Two point correlators $C^{\alpha \alpha}(1, \ell')$ for $\ell'=1, 2$ are then computed via MPS-MPO contraction. The algorithm proceeds iteratively by extending the causal evolution to successive sites $\ell^{\prime} > 2$: at each step, gates with the light cone of site $\ell^{\prime}$ are applied, the corresponding correlator $C^{\alpha \alpha}(1, \ell')$ is evaluated using MPS-MPO contraction, and projective measurement is performed immediately thereafter. This procedure is repeated until the end of the chain is reached, yielding the full spatial profile of equal-time correlators with fixed reference site $\ell = 1$.

This protocol naturally generalizes to the evaluation of equal-time two-point correlation functions with an arbitrary reference site $\ell$. As illustrated in panel (b) of Fig.~\ref{fig:Appendix_Fig1}, we take $\ell = 5$ as a representative example. All sites $\ell' < \ell$ are sequentially time evolved and projectively measured in a chosen basis, thereby disentangling these sites from the system. The two-site unit cell containing the reference site $\ell=5$ (and its neighbor $\ell=6$) is then evolved to the target time without projection, preserving entanglement necessary for correlation evaluation. The correlators $C^{\alpha\alpha}(5, \ell')$ with $\ell'=5, 6$ are computed via MPS-MPO contraction. For $\ell' > 6$, the system is further evolved along the causal light cone, with each correlation $C^{\alpha\alpha}(5, \ell')$ evaluated immediately prior to performing a projective measurement at site $\ell^{\prime}$. Correlators with $\ell' < \ell$ can be accessed by designating $\ell'$ as the reference site and computing $C^{\alpha\alpha}(\ell', \ell)$, or equivalently by reversing the spatial evolution direction. This entangled measurement strategy is fully general and accommodates extensions to higher-order correlation functions as well as mixed-component observables involving distinct operators across different sites.

\begin{figure*}
    \hspace{-0.2in}
    \includegraphics[width=1\textwidth]{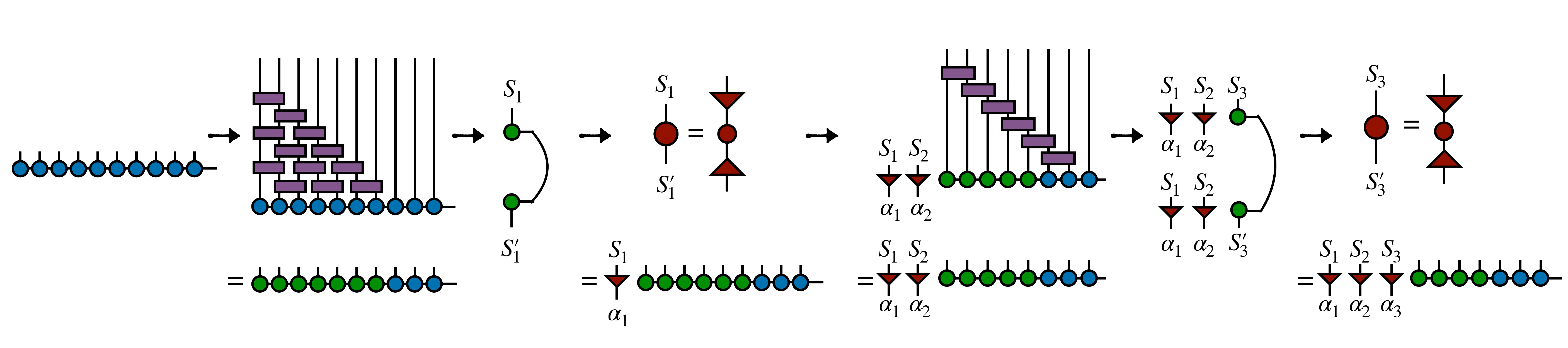}
    \caption{Schematic representation of entangled measurement (EM) via reduced density matrix (RDM) sampling in the SEBD framework. The many-body wavefunction is encoded as an MPS and evolved forward in time by sequentially applying two-site gates within the causal light cones of local unit cells. Once a unit cell reaches its target physical time, it is disentangled via a projective measurement. Local observables $S^{\alpha}_{\ell}$ ($\alpha = x, y, z$) are evaluated using the one-body RDM (1-RDM) at site $\ell$. Projective measurements are implemented by diagonalizing the 1-RDM, selecting an eigenstate with probability given by the corresponding eigenvalue, and collapsing the wavefunction accordingly. A random number $\mathcal{R} \in [0, 1]$ drawn from a uniform distribution is used to sample from the eigenvalue distribution. The diagram illustrates this procedure for the first three sites $\ell = 1 \rightarrow 3$, following time evolution to $t=3$ under the kicked Ising dynamics.}
    \label{fig:Appendix_Fig2}
\end{figure*}

\section{Boosting Sampling Efficiency through RDM-Based Projective Measurements} \label{appendix:RDM_sampling}

Sampling in the SEBD framework is highly adaptable due to full access to many-body wavefunctions through tensor network representations. In the entangled measurement (EM) protocol, beyond evaluating physical observables via MPS-MPO contraction, one can alternatively reduced density matrix (RDM) sampling, wherein local observables are measured in their locally optimal basis. This approach enhances sampling efficiency by aligning measurement axes with the eigenbasis of local density matrices.

The procedure is illustrated schematically in Fig.~\ref{fig:Appendix_Fig2}. For simulations sweeping from left to right, the MPS is brought into canonical form with the orthogonality center positioned on the first site of each unit cell. The corresponding one-body RDM (1-RDM), denoted $\rho^{A}$, is constructed. Local observables $S^{\alpha}_{\ell}$ with $\alpha \in \{x, y, z\}$ are then computed as
\begin{eqnarray}
    \langle S^{\alpha}_{\ell} \rangle = \mathrm{tr} [\rho^{A} S^{\alpha}_{\ell}].
\end{eqnarray}
To execute a projection measurement, the 1-RDM is first diagonalized to obtain its eigenvalues $\lambda_{i}$ and corresponding eigenstates $|\psi_{i}\rangle$. A measurement outcome is then sampled by drawing a random number $\mathcal{R} \in [0, 1]$ and identifying the eigenstate $|\psi_{k}\rangle$ such as $\sum_{i=1}^{k-1} \lambda_{i} < \mathcal{R} \leq \sum_{i=1}^{k} \lambda_{i}$. The MPS is projected onto $|\psi_{k}\rangle$ at the current site, and the post-measurement wavefunction is normalized. The orthogonality center is subsequently shifted to the next site, where the process is repeated. By aligning the measurement basis with the locally optimal basis, RDM sampling can significantly improve statistical efficiency compared to fixed-basis bitstring sampling---particularly in strongly entangled regimes or when estimating observables with small expectation values.

\begin{figure}[h!]
    \hspace{-0.2in}
    \includegraphics[width=0.52\textwidth]{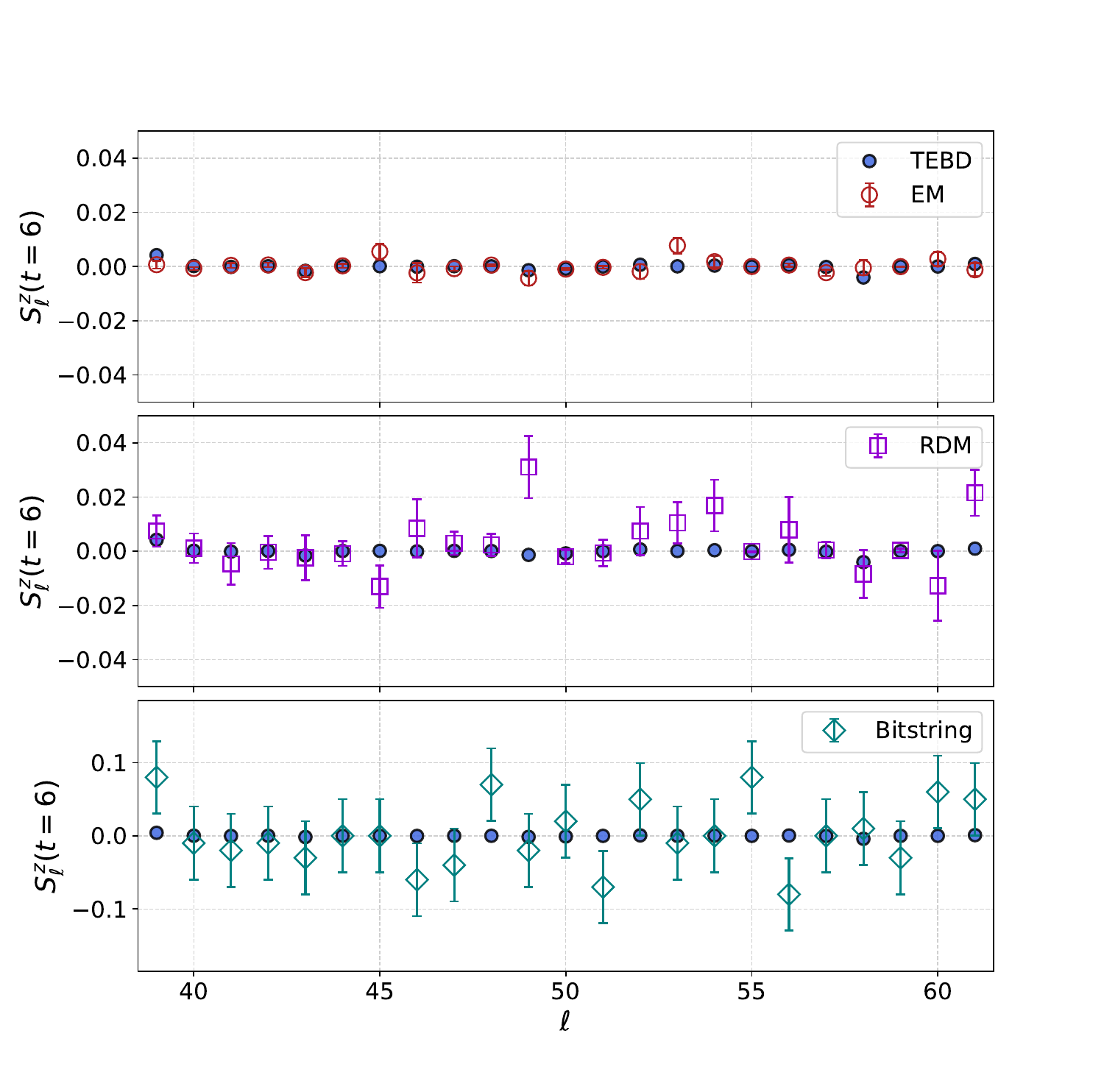}
    \caption{Comparison of sampling accuracy and efficiency across three protocols for evaluating the local observable $S^{x}_{\ell}(t)$. Shown are results from (a) EM using MPS-MPO contraction (red circles), (b) EM via RDM sampling (purple squares), and (c) bitstring sampling based on the Born rule (green diamonds). All methods are benchmarked against reference values obtained by evolving a randomly initialized MPS from $t=0$ to $t=6$ using TEBD. Each data point represents an average over $N_{s}=100$ independent SEBD trajectories on a chain of length $N=100$, with a truncation threshold $\epsilon = 10^{-8}$.}
    \label{fig:Appendix_Fig3}
\end{figure}

We benchmark the sampling efficiency and accuracy of the RDM sampling, compared with MPS-MPO contraction and bitstring samplings. Starting form a randomly initialized MPS on a chain of $N_{s}=32$ sites, the system is evolved to time $t=6$, and local observables $S^{z}_{\ell}(t)$ are evaluated. To ensure a fair comparison. all methods are allocated an equal number of samples, fixed at $N_{s}=100$. As shown in panels (a)-(c) in Fig.~\ref{fig:Appendix_Fig3}, both MPS-MPO contraction and RDM sampling faithfully reproduce the TEBD reference values within one standard error across nearly all sites, despite the modest sample count. MPS-MPO contraction yields slightly reduced deviations compared to RDM sampling. In stark contrast, bitstring sampling exhibits significantly poorer performance: observable estimates display pronounced statistical fluctuations and substantially wider error bars, with standard errors exceeding those of EM-based methods by several factors. To highlight this discrepancy, the vertical axis in Fig.~\ref{fig:Appendix_Fig3}(c) is scaled to twice the range in used in panels (a) and (b), underscoring the amplified statistical noise inherent to bitstring sampling. This degradation is especially severe for observables with small expectation values, where sampling noise dominates due to near-equal probabilities of opposing outcomes. In all cases, the EM protocol maintains uniformly low variance and robust accuracy, demonstrating its superior efficiency for extracting local observables.

\begin{figure}[h]
    \hspace{-0.2in}
    \includegraphics[width=0.48\textwidth]{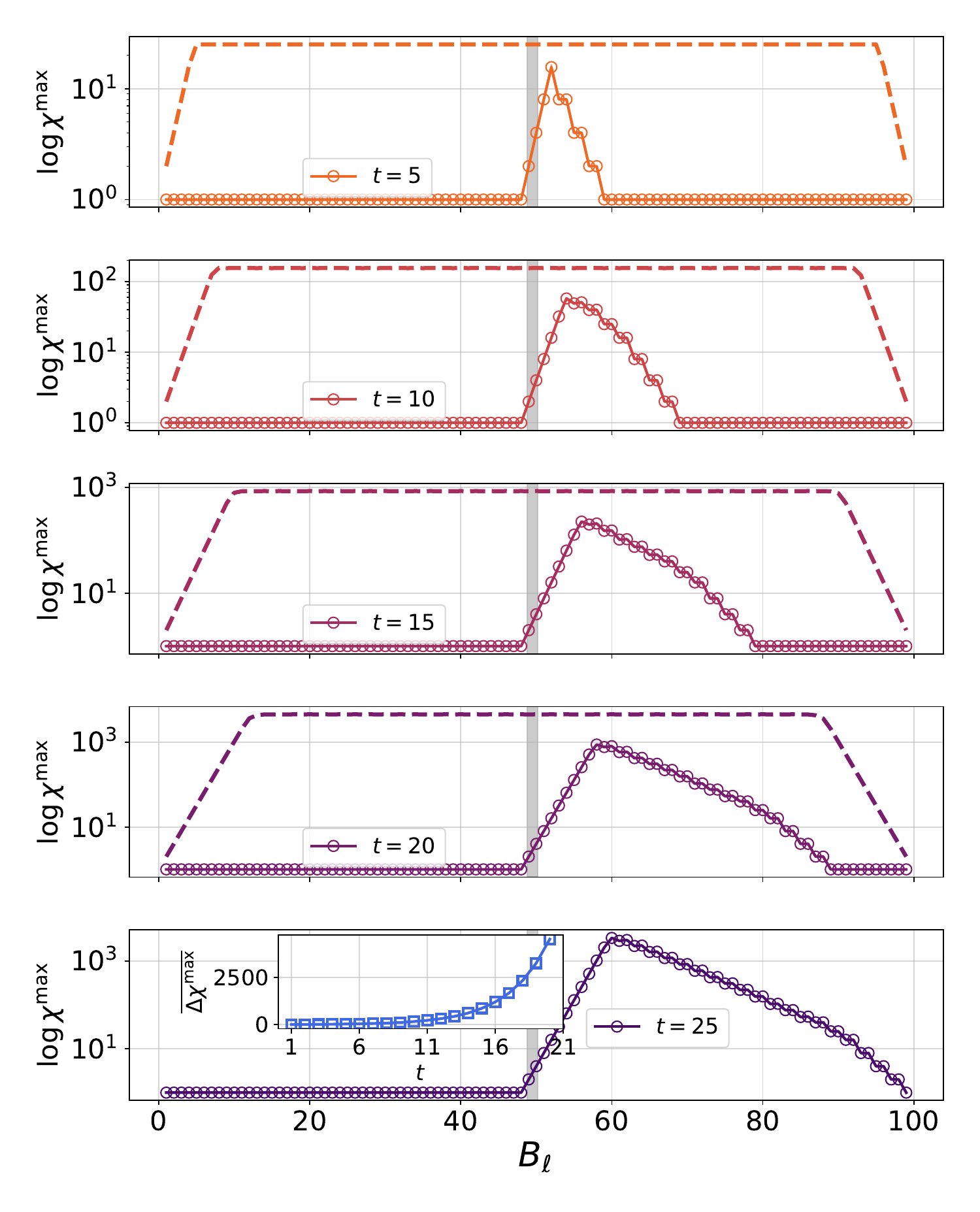}
    \caption{Bond-resolved comparison of the MPS bond dimensions $\chi(B_{\ell})$ between SEBD and TEBD simulations of the kicked Ising model at representative times $t = 5$, $10$, $15$, $20$, and $25$. Simulations are conducted on a 1D chain of $N = 100$ sites initialized in a N\'{e}el state, with a truncation threshold $\epsilon = 10^{-8}$. Panels (a)-(e) show spatial profiles of the bond dimension $\chi(B_{\ell})$, where bond $B_{\ell}$ denotes the bond connecting sites $\ell$ and $\ell + 1$. The shaded region highlights the central bonds $B_{\ell}=49$ and $50$, corresponding to sites $(49, 50)$ and $(50, 51)$, respectively. In SEBD, the unit cell containing sites $\ell=49$ and $50$ is evolved to the target time, measured, and projected out. To ensure a fair comparison, bond dimensions are recorded immediately prior to projection. The inset of panel (e) displays the sample-averaged difference in maximum bond dimension, $\overline{\Delta \chi^{\max}}$, with grows with time and underscores the substantial entanglement suppression---and corresponding computational advantage---achieved by SEBD relative to TEBD.}
    \label{fig:Appendix_Fig4}
\end{figure}

\section{Controlling Bond Dimension Scaling through Intermediate Measurements in SEBD}  \label{appendix:ising_chi} 

The bond dimension $\chi$ of an MPS serves as a central variational parameter that determines both the computational cost and memory requirements of tensor network simulations. It grows approximately as $\chi \sim e^{S_{\mathrm{vN}}}$, where $S_{\mathrm{vN}}$ denotes the bipartite von Neumann entanglement entropy~\cite{Schollwock2011}. Consequently, any reduction in entanglement directly translates to a lowered bond dimension, with substantial algorithmic benefits.

Given the entanglement suppression achieved by the SEBD framework, a significant decrease in $\chi$ relative to conventional TEBD is anticipated. This behavior is confirmed in Fig.~\ref{fig:Appendix_Fig4}, which compares the spatial profile $\chi(B_{\ell})$ between SEBD and TEBD at representative times $t=5$ and $t=25$, immediately prior to applying projective measurements on the central unit cell spanning sites $\ell=49, 50$. At all times, SEBD yields markedly lower bond dimensions across the chain. Importantly, the sample-averaged gap in maximum bond dimension, $\overline{\Delta\chi^{\max}}$, increases from $\sim 100$ at $t=10$ to over $\sim 3600$ at $t=20$, highlighting the exponential computational advantage afforded by light-cone-based evolution combined with intermediate projective measurements. As shown in the inset of Fig.~\ref{fig:Appendix_Fig4}(d), $\overline{\Delta\chi^{\max}}$ exhibits exponential growth in time, in direct correspondence with the linear increase in the entanglement entropy gap $\overline{\Delta S^{\max}_{\mathrm{vN}}}$. This exponential scaling is robust across different truncation thresholds, although the absolute magnitude of $\chi$ depends on the chosen truncation error $\epsilon$. All simulations presented in Fig.~\ref{fig:Appendix_Fig4} were performed with a fixed threshold of $\epsilon=10^{-10}$.

\section{Scaling Efficiency and Sampling Accuracy of SEBD in Continuous-Time Quantum Dynamics}  \label{appendix:heisenberg}

\begin{figure}
    \hspace{-0.2in}
    \includegraphics[width=0.48\textwidth]{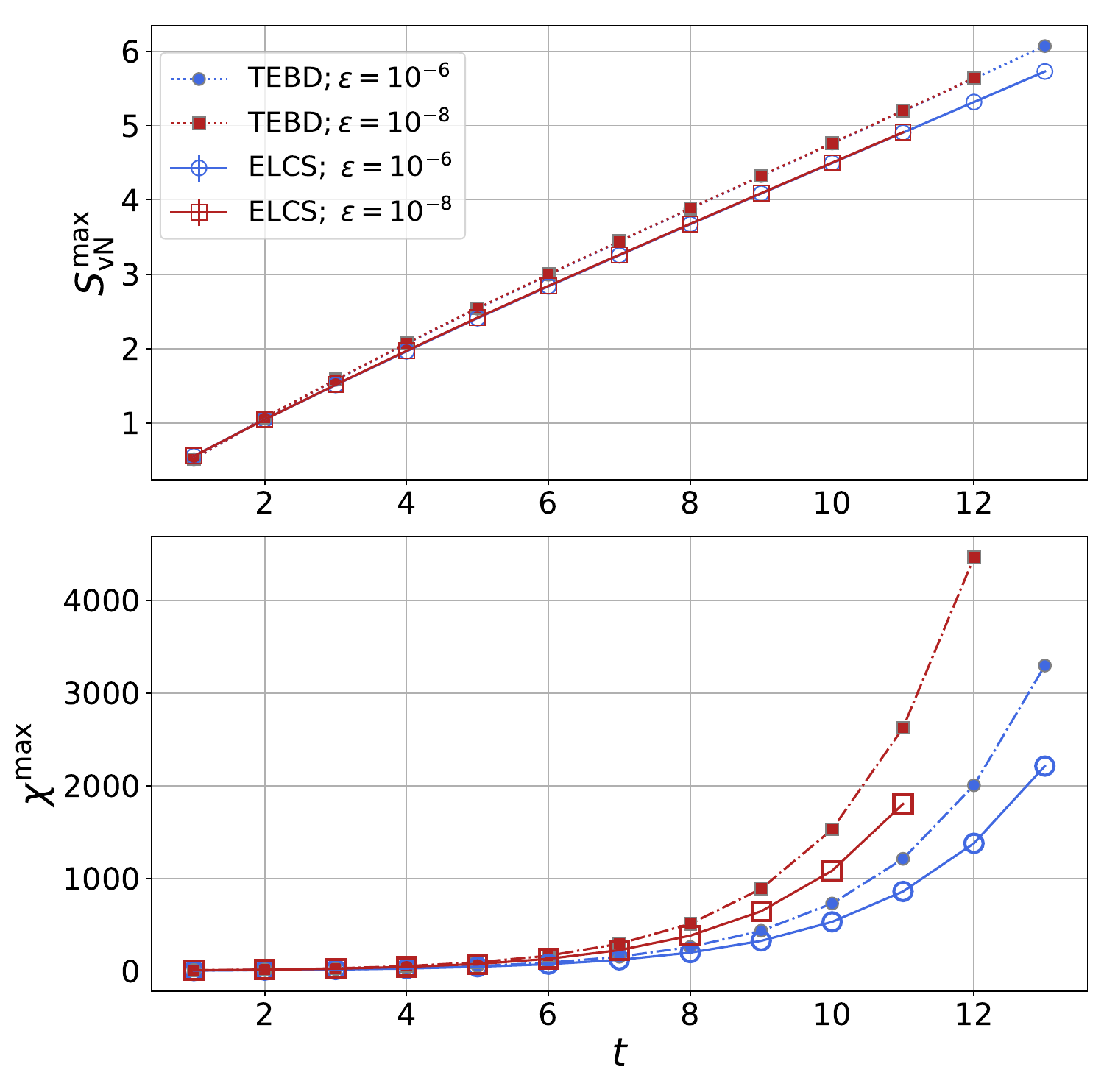}
    \caption{Comparison of sample-averaged (a) maximum von Neumann entanglement entropy $\overline{S^{\max}_{\mathrm{vN}}}$ and (b) maximum bond dimension $\overline{\chi^{\max}}$ as functions of real time for the spin-$1/2$ Heisenberg model, computed using SEBD (open symbols) and TEBD (filled symbols) across multiple truncation thresholds $\epsilon$. SEBD incorporates both causal light-cone propagation and intermediate projective measurements, which leads to a substantial and systematic suppression of entanglement growth and bond dimension relative to TEBD. The discrepancy between the two methods grows steadily with time, demonstrating SEBD's better scabiluty for long-time dynamics. Simulations are performed on chains of $N = 400$ sites for $\epsilon=10^{-6}$ and $N=500$ sites for $\epsilon=10^{-8}$, with a fixed Trotter step size $\Delta\tau = 0.1$.}
    \label{fig:Appendix_Fig5}
\end{figure}

The suppression of entanglement entropy and bond dimension growth in SEBD is a robust, model-independent feature that originates from its integration of projective measurements with the interleaved temporal and spatial evolution. As demonstrated previously for the kicked Ising model, this advantage persists in the spin-$1/2$ Heisenberg chain. Fig.~\ref{fig:Appendix_Fig5}(a) shows the sample-averaged maximum von Neumann entanglement entropy, $\overline{S^{\max}_{\mathrm{vN}}}$, computed using both SEBD and TEBD for a system of $N=400$ sites initialized in a N\'{e}el state. Simulations are performed with truncation thresholds $\epsilon=10^{-6}$ and $10^{-8}$ to confirm numerical robustness. In all cases, SEBD produces markedly lower entanglement than TEBD, reflecting the systematic disentangling effect of intermediate measurements. In the continuous-time limit, linear fits to $\overline{S^{\max}_{\mathrm{vN}}}$ reveal consistently smaller entanglement-growth rate for SEBD, demonstrating its superior capacity to constrain entanglement spreading and thereby reduce the computational complexity of tensor‑network simulations across a broad class of Hamiltonians.

\begin{figure}
    \hspace{-0.25in}
    \includegraphics[width=0.5\textwidth]{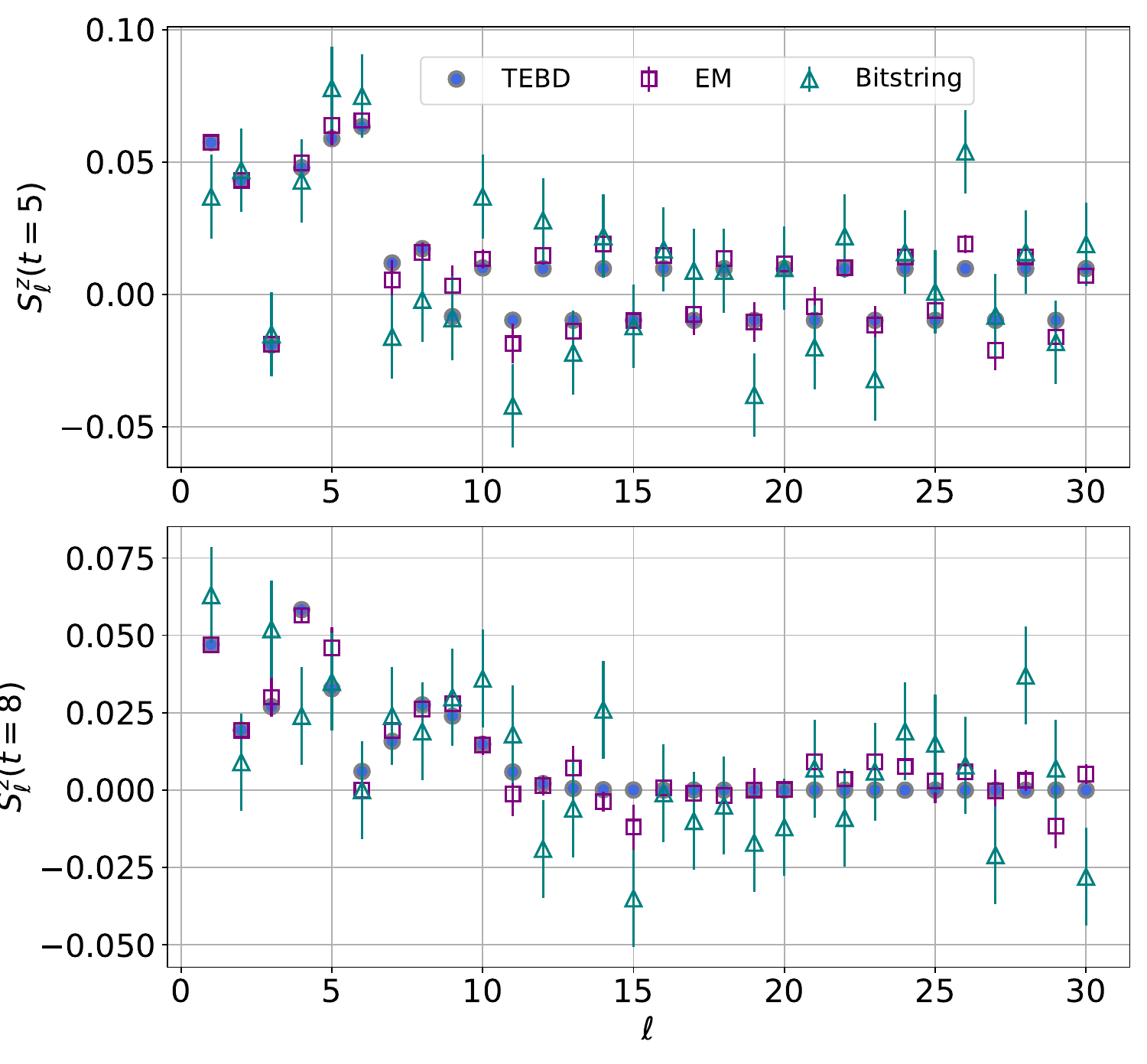}
    \caption{Comparison of sampling efficiency between EM and bitstring sampling protocols for evaluating local observable $S^{z}_{\ell}(t)$ at evolution time (a) $t = 5$ and (b) $t = 8$, respectively. The system is initialized in a N\'{e}el state evolved under the Heisenberg Hamiltonian on a chain of length $N = 400$. Each method uses $N_{s}=1000$ samples and a fixed truncation threshold of $\epsilon=10^{-6}$. While both protocols reproduce the expected time-dependent magnetization profiles, EM achieves markedly reduced statistical fluctuations, especially near regions where $\langle S^{z}_{\ell}(t) \rangle \approx 0$, demonstrating its superior sampling efficiency for extracting local observables.}
    \label{fig:Appendix_Fig6}
\end{figure}

Fig.~\ref{fig:Appendix_Fig5}(b) presents the time evolution of the sample-averaged maximum bond dimension, $\overline{\chi^{\max}}$, obtained using SEBD (open symbols) and TEBD (filled symbols). While both methods exhibit exponential growth in $\overline{\chi^{\max}}$ with time, the growth rate is substantially reduced in SEBD. Consequently, the bond-dimension gap between SEBD and TEBD widens systematically over time, signaling the increasing computational efficiency of SEBD at late evolution time. This behavior is a direct consequence of SEBD's intermediate measurements and confirm its scalability for long-time quantum dynamics.

Because the efficiency gains of EM protocol are independent of the underlying time discretization scheme, similar sampling advantages are expected to persist in the continuous-time limit. To test this hypothesis, we benchmark EM against bitstring sampling in the spin-$1/2$ Heisenberg model. Fig.~\ref{fig:Appendix_Fig6} compares the local magnetization $S^{z}_{\ell}(t)$ at representative times $t=5$ and $t=8$, following real-time evolution from an initial N\'{e}el state on a chain of $N=400$ sites. For clarity, results are shown for the first 30 sites ($\ell = 1 \rightarrow 30$). At fixed sample size $N_{s}=1000$, EM closely reproduces the TEBD reference values across nearly all sites, with uncertainties confined within one standard error. In contrast, bitstring sampling exhibits large fluctuations and significantly broader error bars, particularly at sites where the expectation values are close to zero. In such regions, the inefficiency of bitstring sampling arises from equal-weight sampling from both eigenstates. These results confirm that EM retains its superior sampling precision and statistical efficiency in the continuous-time regime, making it broadly applicable across discrete and continuous dynamics settings.

\newpage

\clearpage
\bibliography{main}

\end{document}